%% file: main.tex
\RequirePackage[2024-06-01]{latexrelease}
\documentclass[twocolumn]{aastex631}

\usepackage{placeins}
\usepackage{xcolor}
\usepackage{amsmath}
\usepackage{amssymb}
\usepackage{chemformula}
\usepackage{comment}

\usepackage{chngpage}
\usepackage{booktabs}
\usepackage{multirow}
\usepackage{array}
\makeatletter
\let\switch@array\relax
\makeatother
\newcolumntype{P}[1]{p{#1}}
\usepackage{makecell}
\usepackage[caption=false]{subfig}
\usepackage{textcomp}

\newcommand\standardstate{{\circ\kern-0.495em-}}

\defcitealias{madhusudhan_new_2025}{M25}
\defcitealias{welbanks_challenges_2026}{W26}
\defcitealias{taylor_are_2025}{T25}
\defcitealias{madhusudhan_carbon-bearing_2023}{M23}

\begin{document}

\title{A Systematic Search for Trace Molecules in the Atmosphere of Exoplanet K2-18~b}

\author{Lorenzo Pica-Ciamarra}
\affiliation{Institute of Astronomy, University of Cambridge,
Madingley Road, Cambridge CB3 0HA, UK}

\author{Nikku Madhusudhan}
\affiliation{Institute of Astronomy, University of Cambridge,
Madingley Road, Cambridge CB3 0HA, UK}

\author{Gregory J. Cooke}
\affiliation{Institute of Astronomy, University of Cambridge,
Madingley Road, Cambridge CB3 0HA, UK}

\author{Savvas Constantinou}
\affiliation{Institute of Astronomy, University of Cambridge,
Madingley Road, Cambridge CB3 0HA, UK} 

\author{Martin Binet}
\affiliation{Institute of Astronomy, University of Cambridge,
Madingley Road, Cambridge CB3 0HA, UK}

\correspondingauthor{Nikku Madhusudhan}
\email{nmadhu@ast.cam.ac.uk} 

\begin{abstract}
The first transmission spectrum of the habitable-zone sub-Neptune K2-18~b with JWST has opened a new avenue for atmospheric characterisation of temperate low-mass exoplanets. The observations led to inferences of methane and carbon dioxide, as well as of dimethyl sulfide (DMS) and/or dimethyl disulfide (DMDS), both potential biosignatures. In the present work we conduct a broad and agnostic search for other chemical species in the atmosphere of K2-18~b. Our exploration includes 661 molecules, spanning a wide range of trace gases, including biotic, abiotic, and anthropogenic gases on Earth. We investigate possible preference for any of these gases, compared to a model only including the previously-detected CH$_4$ and CO$_2$, using three metrics: (a) preference in the JWST mid-infrared (MIR) spectrum, (b) preference in the JWST near-infrared spectrum, for species preferred in MIR and (c) plausible sources of production. We find that only DMS consistently results in Bayes factors $\ln B \geq 2.0$ across the datasets considered independently, though in the near-infrared this depends on detector offsets, as previously reported. The threshold of $\ln B \geq 2.0$ is motivated by the conventional threshold of $\ln B \geq 2.5$ for moderate preference, allowing for an empirical uncertainty of 0.5. A few other gases also provide comparable fits to a subset of the data or only with some of the retrieval codes used, but with limited known plausible sources. Our study highlights the need for further observations to distinguish between possible trace gases in the atmosphere of K2-18~b and theoretical work to establish their plausible sources.
\end{abstract}

\section{Introduction} \label{sec:intro}
The atmospheric characterisation of temperate ($T_{\rm eq} \lesssim 400 {\rm~K}$) planets is the new frontier of exoplanet science. Sub-Neptune planets are the largest class of currently known exoplanets \citep[e.g.,][]{fulton_california-kepler_2018}, but they have no analogue in the solar system, such that any empirical insights into their nature must come from remote atmospheric observations. The start of JWST operations has led to a revolution in our ability to characterise temperate sub-Neptune atmospheres. JWST near-infrared (NIR) observations of the habitable-zone planet K2-18~b have led to the first detections of carbon-bearing molecules (CO$_2$ and CH$_4$) in a temperate sub-Neptune \citep{madhusudhan_carbon-bearing_2023}, with tentative evidence for dimethyl sulfide (DMS), a possible biosignature gas \citep{Pilcher_Bio_2003, Domagal-Goldman_2011, seager_biosignature_2013}. \citet{schmidt_comprehensive_2025} disputed this finding, claiming no evidence for CO$_2$ and DMS. However, a later analysis by \citet{hu_water-rich_2025}, which considered additional NIR observations, confirmed the initial results presented in \citetalias{madhusudhan_carbon-bearing_2023}. Similar results to K2-18~b were obtained for the temperate sub-Neptune TOI-270~d, with the analysis of JWST near-infrared spectra 
revealing CH$_4$ and CO$_2$, as well as tentative inferences of CS$_2$ and H$_2$O \citep{holmberg_possible_2024, benneke_jwst_2024}. 

More recently, the first observations have been reported in the mid-infrared for a temperate sub-Neptune, \citep[][hereafter M25]{madhusudhan_new_2025} with the JWST MIRI spectrum of K2-18~b. They reported evidence for the presence of DMS and/or DMDS, both possible biosignature gases \citep{Pilcher_Bio_2003, Domagal-Goldman_2011}, at a 3$\sigma$ significance compared to a model including CH$_4$ and CO$_2$ only. However, each of these two molecules individually is only tentatively inferred due to the degeneracy between their spectral features \citep{madhusudhan_new_2025}, requiring further observations for a more robust assessment. In particular, \citet{luque_insufficient_2025}, performing a joint analysis of the data, found that the results are dominated by the near-infrared spectrum, and the evidence for presence of DMS depends on the adopted data reduction pipeline, ranging between no evidence and moderate ($\ln B = 2.8$) evidence, consistent with or higher than the corresponding retrieval in \citetalias{madhusudhan_carbon-bearing_2023} with the NIR data alone.   New observations in the near-infrared presented in \citet{hu_water-rich_2025} can also provide, depending on the model used, up to $\ln B = 3.5$ preference for a combination of DMS, N$_2$O, and CH$_3$SH over a model only including six standard CNO species (CH$_4$, CO$_2$, CO, NH$_3$, HCN, H$_2$O). These observations open a new avenue for studying habitable-zone exoplanets, including the search for biosignatures in their atmospheres. 

The MIRI observations of K2-18~b are at the frontier of the capabilities of JWST and test the limits of its sensitivity. As such, they represent the most advanced analysis possible for habitable-zone exoplanets with present facilities. However, given their low signal-to-noise and low resolution, these datasets also open new challenges. For example, as evidenced by the low significance of the DMS and DMDS inferences (\citetalias{madhusudhan_new_2025}, \citealp{luque_insufficient_2025}, \citealp{hu_water-rich_2025}), follow-up efforts are needed both to increase the quality of the observations as well as to robustly interpret them. Efforts in this direction (\citealp[][hereafter T25]{taylor_are_2025}; \citealp[][hereafter W26]{welbanks_challenges_2026}) have explored agnostic approaches to find alternative interpretations for the existing data. While both the \citetalias{taylor_are_2025} and the \citetalias{welbanks_challenges_2026} approaches are agnostic, they rely on different levels of model complexity and physical plausibility. On the one hand, \citetalias{taylor_are_2025} consider parametric Gaussian models to fit the MIRI transmission spectrum, finding up to $\sim $2$\sigma$ Bayesian preference for a model with features compared to a flat line. While the negative Gaussian models reported in \citetalias{taylor_are_2025} are unphysical, the study nevertheless serves the purpose of independently verifying the potential for spectral features in the data. 

\begin{figure*}[t!]
	\centering
	\includegraphics[width=1\textwidth]{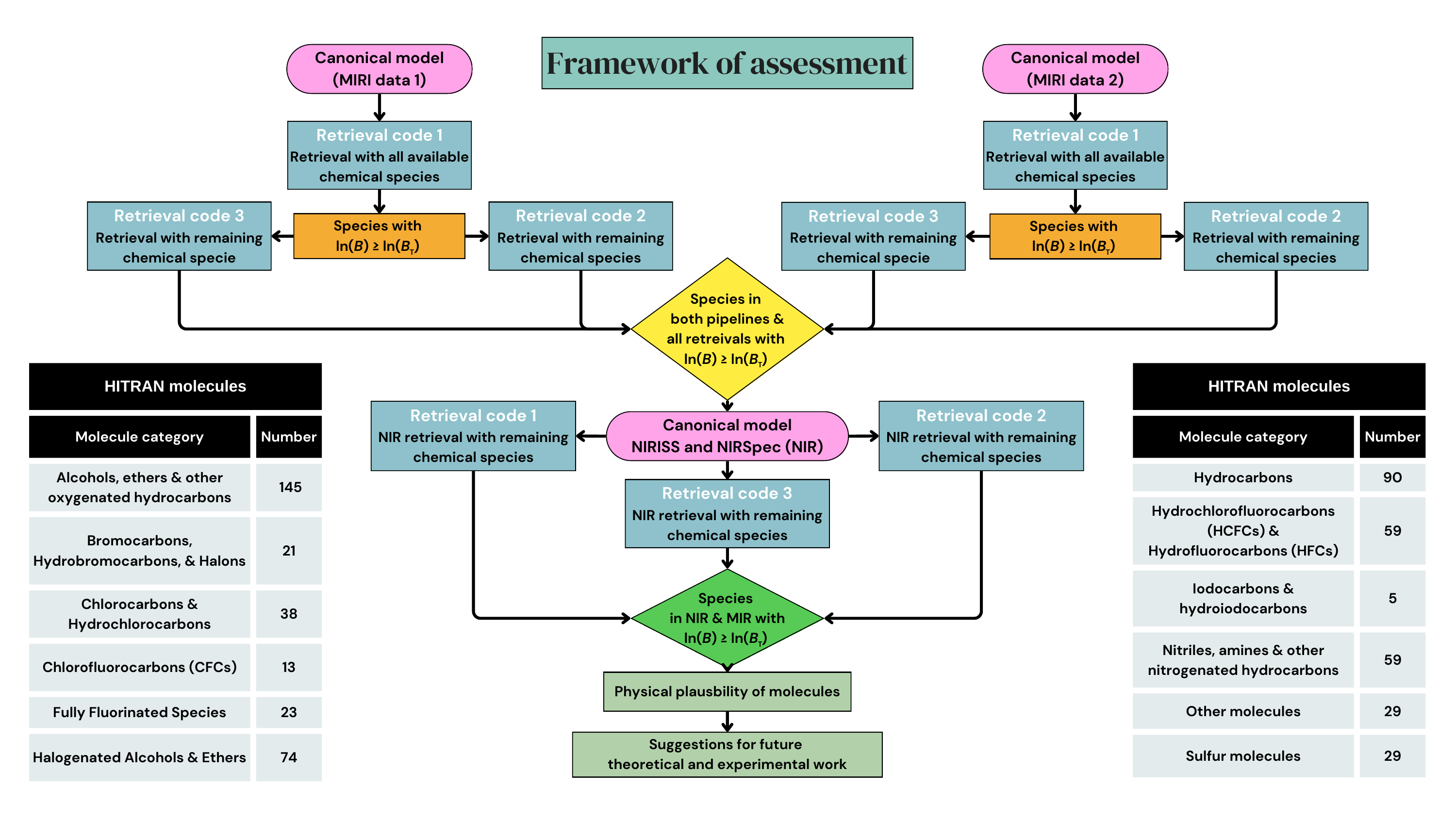}
    \caption{The flow chart shows our framework of assessment for possible molecules in the atmosphere of K2-18~b. Based on the inference of carbon dioxide (CO$_2$) and methane (CH$_4$) in the atmosphere of K2-18~b \citep{madhusudhan_carbon-bearing_2023}, retrievals adopting canonical models are carried out following \citetalias{madhusudhan_new_2025}, assuming hydrogen dominated atmospheres and including the mixing ratios of CO$_2$, CH$_4$, and X as free parameters, with X in this work being any of 661 species explored. The evidence for X is then computed by performing a Bayesian comparison between the relevant canonical model and a nested baseline model, with only X removed. Several independent retrieval codes can be used to increase the reliability of such a framework. In this work, retrieval codes 1, 2, and 3 are \texttt{POSEIDON}, \texttt{VIRA}, and \texttt{petitRADTRANS}, respectively. Retrievals were initially performed with the \texttt{POSEIDON} retrieval code on the two available MIRI datasets for  K2-18~b, resulting from the \texttt{JExoRES} and \texttt{JexoPipe} data reduction pipelines (MIRI data 1 and 2, in the pink curved boxes at the top left and top right). Nearly all molecules in the HITRAN cross-section database were used in these initial retrievals, as shown in the tables on the lower left and lower right hand side of the figure. Additionally, some other available molecules, as well as atoms and ions, were considered, where originally included in \texttt{POSEIDON}. A threshold Bayes factor $\ln B_T$ is chosen, and chemical species for which $\ln B\geq \ln B_T$ is found are identified (orange rectangles), and follow-up retrievals are performed with \texttt{VIRA} and \texttt{petitRADTRANS}. Chemical species for which $\ln B\geq \ln B_T$ is found across both reduction pipelines and all three retrieval frameworks (yellow diamond) are passed to the next step, where we include them in atmospheric retrievals on the near-infrared (NIR) data from the \texttt{NIRISS} and \texttt{NIRSpec} JWST transits (pink curved box in the middle) of K2-18~b \citep{madhusudhan_carbon-bearing_2023}. Finally, retrievals with \texttt{VIRA}, \texttt{petitRADTRANS}, and \texttt{POSEIDON} are used to find if there are species for which $\ln B\geq \ln B_T$ can be reached in the NIR data (green diamond). We then discuss the physical plausibility of promising species being present in the atmosphere of K2-18~b. In this work, we chose $\ln B_T = 2.0$. This is informed by the threshold $\ln B = 2.5$ for moderate evidence proposed by \citet{trotta_bayes_2008}, and accounting for possible systematic uncertainties of up to $\pm 0.5$ on the estimation of $\ln B$, as discussed in Section \ref{sec:species_selection}.} 
    \label{Flow chart figure}
\end{figure*}

On the other hand, \citetalias{welbanks_challenges_2026} pursue a different approach by exploring a larger suite of molecules than the 20 explored in \citetalias{madhusudhan_new_2025}. At the outset, \citetalias{welbanks_challenges_2026} reproduce a $\sim$3$\sigma$ preference for models including DMS and/or DMDS in addition to CO$_2$ and CH$_4$, relative to a baseline model with only CH$_4$ and CO$_2$ included, as reported in \citetalias{madhusudhan_new_2025}. In addition, \citetalias{welbanks_challenges_2026} also find three other molecules with $\ln B \geq 2.5$ preference (equivalent to 2.8$\sigma$ significance, when using the inverse \citet{sellke_calibration_2001} and \citet{trotta_bayes_2008} conversion), corresponding to the lower threshold for moderate evidence \citep{trotta_bayes_2008}, over the baseline model. For these cases, they find significances between 2.8-3.1 $\sigma$, which are comparable to their equivalent significances of 2.7-3.0 $\sigma$ for DMS, DMDS or DMS+DMDS, considering a typical error of 0.1-0.2 $\sigma$ in such calculations \citepalias{madhusudhan_new_2025}. 
The exploration of \citetalias{welbanks_challenges_2026} focused only on the MIRI data reported in \citetalias{madhusudhan_new_2025}. Therefore, open questions remain on the robustness of their findings when the near-infrared data reported in \citet{madhusudhan_carbon-bearing_2023} are considered.

As discussed in \citetalias{madhusudhan_new_2025}, such agnostic explorations are important to establish the space of possible explanations to the data. However, the physical plausibility of most of the 90 hydrocarbons considered in \citetalias{welbanks_challenges_2026} for an exoplanetary atmosphere remains unclear. For example, among the three gases they find above the threshold for moderate model preference, only propyne was discussed as potentially plausible in the atmosphere of K2-18~b. The remaining gases are produced in small quantities on Earth, primarily biotic or anthropogenic, with few sources known in other planetary environments. We discuss the physical plausibility of these gases in section~\ref{sec:plausibility-others}. Furthermore, while \citetalias{welbanks_challenges_2026} focused on hydrocarbons, the much larger space of molecules involving other prominent elements, including N and S, remains unexplored.

A reliable inference of a chemical signature in an exoplanetary atmosphere requires consideration of all the available data and an assessment of its feasibility in the given context. In the present work, we report a systematic and wide-ranging search for trace molecules in the atmosphere of K2-18~b using all available transmission spectra with JWST reported in \citet{madhusudhan_carbon-bearing_2023} and \citetalias{madhusudhan_new_2025}. We conduct in-depth and agnostic atmospheric retrievals exploring a large ensemble of 661 molecules, spanning molecules with abiotic, biotic and anthropogenic sources on Earth. We assess the molecules based on multiple lines of evidence, across the available datasets and multiple retrieval frameworks, and their physical plausibility in the context of K2-18~b. 

In what follows, we begin, in Section \ref{sec:assess}, with a discussion of prevalent practices with regards to model selection, standards of evidence, and biosignature assessment, in the characterisation of exoplanetary atmospheres. Next, in Section \ref{sec:methods}, we describe the protocol for our systematic search and the retrieval approach. We present our results in Section \ref{sec:results} and discuss their physical plausibility in Section \ref{Physical plausibility of results section}. We summarise our work and discuss future directions in Section \ref{sec:discussion}.  

\section{Considerations from Previous Work} \label{sec:assess}

The launch of JWST \citep{gardner_james_2006} has opened a new era in the characterisation of exoplanetary atmospheres. The quality of available data requires new efforts in establishing appropriate methods for their analysis. The spectroscopic capability of JWST instruments spans a wide range, from higher resolution NIRISS (R$\sim$700 for NIRISS SOSS) and NIRSpec (R$\sim$2700 with high-resolution gratings) spectroscopy in the near-infrared to low-resolution spectroscopy and photometry with MIRI (R$\sim$100 for MIRI Low Resolution Spectroscopy (LRS)) in the mid-infrared. The presently available MIRI LRS spectra, in particular, can sometimes be comparable in data quality to the best Hubble Space Telescope (HST) spectra of exoplanets obtained in the pre-JWST era, with significantly lower observing time and much broader wavelength range. It is thus pertinent to review prevalent practices on the analysis and interpretation of such data in order to inform the next steps towards the characterisation of exoplanetary atmospheres with JWST. 

\subsection{Model Selection}

A central consideration in atmospheric retrievals of exoplanets is the complexity and dimensionality of the model considered given the data at hand. Atmospheric retrievals of low-mass exoplanets in the pre-JWST era provided initial insights into the model requirements. 
In the case of the sub-Neptune K2-18~b, early retrievals were conducted using a low-resolution spectrum obtained with HST/WFC3 \citep{benneke_water_2019, tsiaras_water_2019, madhusudhan_interior_2020}. The nominal model in \citet{benneke_water_2019} included as free parameters the mixing ratios for seven chemical species, the cloud-top pressure for a gray cloud deck, and an isothermal $P$-$T$ profile, for a total of nine free parameters.  On the other hand, the largest model in \citet{tsiaras_water_2019} considered free mixing ratios for five species, and their nominal model included only two species. As another example, \citet{madhusudhan_interior_2020} included free mixing ratios for five species, as well as a nonisothermal $P$-$T$ profile and inhomogeneous clouds, for a total of 16 free parameters.

With the advent of JWST, unprecedented data quality was obtained in the near-infrared, which led to significantly more complex models for atmospheric retrievals. For example, \citet{madhusudhan_carbon-bearing_2023} considered the mixing ratios of 11 species as free parameters when performing atmospheric retrievals on the atmosphere of K2-18~b. However, the presently available data obtained from the mid-infrared instrument, MIRI LRS, is of comparable quality to the near-infrared HST WFC3 data, albeit at a fraction of the observing time. As model complexity (e.g., the number of free parameters included in the model) needs to match data quality (e.g., the spectral resolution and signal-to-noise of available data), retrieval set-ups used for the analysis of comparable HST WFC3 data act as a good guideline for the appropriate choices when analysing JWST MIRI LRS data. This has indeed been followed in the atmospheric characterisation of hot Jupiter WASP-17~b with a dataset including both JWST MIRI and HST WFC3 data, which used models including 4-7 molecules \citep{grant_jwst-tst_2023}. An even simpler set-up was adopted in several of the models considered by \citet{powell_sulfur_2024}, which only included two species, H$_2$O and SO$_2$. 

In the above mentioned cases, the selection of species to be included in the atmospheric retrievals was broadly based on thermochemical and/or photochemical expectations for the planets being studied. The analysis of MIRI LRS data of K2-18~b by \citetalias{madhusudhan_new_2025}, working with data of similar quality to those analysed in the studies mentioned above, attempted to formalise the model selection process. In order to do so, \citetalias{madhusudhan_new_2025} introduced a new approach which is also more conducive to the exploration of a larger parameter space.

\citetalias{madhusudhan_new_2025} started by considering a ``maximal'' model, which included 20 species, the largest set at the time for JWST observations of sub-Neptunes. Inspection of the posterior distributions revealed that only one species, DMDS, showed a clear peak. In order to identify other potential contributors to the spectrum, DMDS was removed from the model. This led to DMS showing a peak in the posterior distribution, upon the removal of which no other molecules stood out. A simpler model with four species, including CH$_4$ and CO$_2$, inferred in the near infrared by \citet{madhusudhan_carbon-bearing_2023}, and DMS and DMDS, was then considered. This canonical model was found to be preferred at a $2\sigma$ level over the maximal model. Following Occam's razor, this simpler case was adopted as the canonical model, against which to compute model preferences; for a discussion on model complexity in a Bayesian setting, see \citet{trotta_bayes_2008}. This canonical model is of similar complexity, and includes a similar number of molecules, to what was used in \citet{madhusudhan_interior_2020} and \citet{grant_jwst-tst_2023}, for comparable data quality. In addition, \citetalias{madhusudhan_new_2025} also noted the strong degeneracy between DMS and DMDS in the MIRI band and, therefore, explored canonical models with only one of DMS or DMDS included besides CH$_4$ and CO$_2$. 

Later, \citetalias{welbanks_challenges_2026} investigated the \citetalias{madhusudhan_new_2025} observations through a new maximal model, with 21 species and 33 free parameters, and also through an array of 93 canonical models. These models were similar to the \citetalias{madhusudhan_new_2025} canonical model, in that each included three chemical species: CH$_4$, CO$_2$, and each of 92 species, one-by-one, to assess whether the MIRI data provided any evidence for their presence. The 92 species included DMS and DMDS, and \citetalias{welbanks_challenges_2026} also considered a case where both DMS and DMDS were included, corresponding to a similar case used by \citetalias{madhusudhan_new_2025}. They successfully replicated the \citetalias{madhusudhan_new_2025} findings that DMS, DMDS, and their combination are preferred at $\sim$3$\sigma$ over a model only including CH$_4$ and CO$_2$. We note that this is to be expected, given that the \texttt{Aurora} retrieval framework used in \citetalias{welbanks_challenges_2026} belongs to the same family as the \texttt{AURA} retrieval code used in \citetalias{madhusudhan_new_2025}. In addition, however, they found three other molecules, not previously explored by \citetalias{madhusudhan_new_2025}, for which the data provided at least moderate evidence ($\geq 2.7\sigma$ according to Jeffreys' scale, see, e.g., \citealp{trotta_bayes_2008}).

Overall, the model complexity in atmospheric retrievals needs to be motivated by the data quality. As shown in \citetalias{madhusudhan_new_2025}, while a maximal model with 20 species is useful for the exploration of a wide parameter space, it may be less favoured in a Bayesian sense than a model whose complexity better matches the data quality, e.g., a nested model including only the few most relevant molecules. Furthermore, for a large exploration of chemical species, e.g., hundreds of species, a retrieval including all the molecules at once is computationally infeasible currently. Choices of maximal models are thus necessarily arbitrary. For example, while \citetalias{madhusudhan_new_2025} included 20 relevant species, future studies could include other plausible molecules, such as isoprene \citep{zhan_assessment_2021}, propyne \citep{huang__probing_2024}, or benzene \citep{Tsai_2024_Sulfur}. Therefore, the approach of selecting a minimal canonical model informed by theoretical and empirical considerations serves a good starting point at the present time, although it is important to ensure that these are not exceedingly simple \citep[e.g.][]{welbanks_degeneracies_2019, welbanks_atmospheric_2022}. Such minimal models, in turn, can be used for exploring a wide array of species sequentially, thus identifying promising candidate species for further investigation, as pursued in \citetalias{madhusudhan_new_2025} and \citetalias{welbanks_challenges_2026}.

\subsection{Detection Significances}
The question of what degree of statistical significance is to be considered enough to claim a discovery is a question every scientific discipline has had to address. The standards vary widely across fields: in biological and medical research, for example, a result with a p-value of $p < 0.05$, corresponding to a $2\sigma$ result, is often considered acceptable, despite doubts as to the appropriateness of this threshold \citep{fisher_statistical_1934, amrhein_earth_2017}. On the other hand, in experimental high energy physics, significances of $ \geq 5\sigma$ are required to claim a discovery \citep{particle_data_group_review_2022}. 

Until recently, the consensus in the field of exoplanet atmospheres seemed to be that the threshold for a chemical detection lied somewhere between $\sim$2-3 $\sigma$, as determined by converting the relevant Bayesian model preference \citep[see, e.g.,][]{trotta_bayes_2008, benneke_how_2013}. For example, \citet{grant_jwst-tst_2023} claimed a detection of quartz clouds in the atmosphere of hot Jupiter WASP-17~b through a JWST MIRI spectrum, when the model with quartz clouds was preferred at $2.6\sigma$ over a (non-nested) model only including a generic aerosol parametrisation. Similarly, \citet{powell_sulfur_2024} considered SO$_2$ to have been detected in the MIRI spectrum of WASP-39~b with significance values varying between $2.5\sigma$ and $4.21\sigma$, depending on the choice of retrieval code and data reduction pipeline. More recently, \citet[][]{liu_unveiling_2025} referred to both a $3.09\sigma$ preference for H$_2$O and a $1.90\sigma$ preference for a gray cloud deck in the hot jupiter HAT-P-14~b as detections. Similarly, \citet{murphy_hst_2025} reported a detection of H$_2$O in the atmosphere of hot Neptune HD~219666~b, based on a model preference of $2.9\sigma$. In all these cases, as discussed above, the number of molecules considered in the models ranged between 4-7 molecules. We highlight that, in all cases discussed, detection significances were obtained by Bayesian model comparison, which provides a metric for the preference of a model compared to a reference model.

The recent inferences of chemical signatures in the sub-Neptune K2-18~b (\citealp{madhusudhan_carbon-bearing_2023}, \citetalias{madhusudhan_new_2025}) have motivated a reconsideration of what statistical significance constitutes a reliable detection. For example, \citet{schmidt_comprehensive_2025} advocate for adopting a higher significance to claim detections of chemical species in exoplanet atmospheres. This represents a shift from the standard employed in earlier works using similar retrieval frameworks, such as \citet{grant_jwst-tst_2023} and \citet{murphy_hst_2025}. A reasonable approach would be to follow the conventional metric of Jeffreys' scale in Bayesian inference of considering a significance between 2.7$\sigma$ and 3.6$\sigma$ as moderate evidence, and at and above 3.6$\sigma$ as strong evidence \citep[e.g.,][]{trotta_bayes_2008}. However, Bayesian model selection itself is not free from pitfalls, and caution needs to be exercised when applying it (\citealp[e.g.,][]{efstathiou_limitations_2008, welbanks_application_2023, nixon_methods_2024}; \citealp{dittmann_notes_2024, thorngren_bayesian_2026}). For any approach chosen, it is also important to use the same metric across all chemical inferences to ensure a uniform standard in the field. 

\subsection{Biosignatures in Context}

A central question in the field is about what molecules at what abundances can be regarded as robust biosignatures, i.e., unambiguous indicators of the presence of life. Various studies in the past have suggested criteria to conclusively identify a biosignature \citep[e.g.][]{SeagerTowards2016, catling_exoplanet_2018, Meadows2022Biosignature}. Most recently, \cite{seager_prospects_2025} reiterated three criteria that need to be met for a detected molecule to be considered as a robust biosignature: (a) it has a distinctive spectral feature versus the dominant gas, (b) it has a plausible production rate, and (c) it has no known significant false positives in the given planetary context.

A number of molecules produced primarily by life on Earth have been suggested in the past as reliable biosignatures for habitable exoplanets. These include molecules such as N$_2$O, DMS, and CH$_3$Cl \citep[e.g.][]{Domagal-Goldman_2011, seager_biosignature_2013, catling_exoplanet_2018, Leung2022Methylated}. The identification of a biosignature would be dependent on the specific context and environment, and rely on multiple lines of evidence \citep{Meadows2022Biosignature}. However, what qualifies as a relevant abiotic source for a potential biosignature molecule remains an open question. Previous assessments of abiotic false positives included possible sources from geological or atmospheric processes, e.g. through photochemistry. For example, while CH$_4$ is predominantly produced by life on Earth, it is also produced in small quantities by geochemical sources and can be produced through atmospheric chemistry in H$_2$-rich environments \citep{schwieterman_exoplanet_2024}. On the other hand, it is also important to ascertain that any proposed abiotic mechanism for an observed molecule is also physically plausible in the context of a given planetary atmosphere, and is consistent with the observed constraints. 

The discussion around the molecule DMS in the habitable-zone sub-Neptune K2-18~b is a case in point. For two decades, DMS has been regarded as a robust biosignature across different environments, in both Earth-like and H$_2$-rich atmospheres \citep{Pilcher_Bio_2003, Domagal-Goldman_2011, SeagerTowards2016, catling_exoplanet_2018, madhusudhan_habitability_2021}. However, the first inference of DMS in K2-18~b (\citealp{madhusudhan_carbon-bearing_2023} and \citetalias{madhusudhan_new_2025}) has prompted a reconsideration of its candidacy as a robust biosignature \citep{seager_prospects_2025}. Recently, \cite{seager_prospects_2025} argued that DMS is no longer a robust biosignature due to potential abiotic formation scenarios (see their Table 1 and Table S2). \citet{seager_prospects_2025} supported this statement by highlighting the abiotic synthesis of DMS in laboratory photochemical experiments \citep{Raulin1975, Reed2024}, laboratory,  and its identification in a comet \citep{Hanni2024} and the interstellar medium (ISM) \citep{2025ApJ...980L..37S}. However, as discussed in \citetalias{madhusudhan_new_2025} (their Section 4.2), neither the laboratory experiments nor the comet detection are able to explain the high inferred abundance of DMS (10-1000 ppm) in the context of the observed atmosphere of K2-18~b. Generally, the presence of a molecule on a comet or in the ISM does not by itself constitute a reliable false positive, given their vastly different environments compared to a dense planetary atmosphere, and the implausibility of volatile delivery and stability through cometary impacts \citep{Leung2022Methylated, madhusudhan_new_2025}. A large number of complex organic compounds, including amino acids \citep{Belloche2013ISM, Inter2022Organic}, which can be produced and remain stable in the cold and low-density environments of the ISM and comets \citep{LeRoy2015Comet} would be unlikely to remain stable and abundant in a planetary atmosphere.  Finally, the closest experiment to the currently evaluated context produced DMS mixing ratios of 0.05 ppmv \citep{Reed2024}, at least 200 times lower than (or 0.5\% of) the inferred DMS abundance and requiring the presence of H$_2$S, which has not been detected.

Furthermore, any criterion for the identification of a biosignature or a false positive needs to be applied consistently. For example, \citet{seager_prospects_2025} identify the following molecules as having no known significant false positives in a terrestrial context: CH$_3$OH, PH$_3$, NH$_3$, and potentially CH$_3$Cl, CH$_3$Br, and N$_2$O. However, other studies have shown that NH$_3$, CH$_3$OH, and CH$_3$Cl are also present in comets \citep{LeRoy2015Comet, Fayolle2017Comet}, including the same comet where DMS was reported \citep{Hanni2024, hanni_nitrogen-_2025}, and PH$_3$ is present in the atmospheres of Jupiter and Saturn \citep{barshay1978chemical, courtin1984composition, orton2000vertical, taylor2004composition}. It may be reasonable to suggest NH$_3$ \citep[e.g.,][]{huang_assessment_2022} or PH$_3$ \citep[e.g.,][]{sousa-silva_phosphine_2020} as biosignatures for terrestrial exoplanets, like Earth and Venus, which are very different environments from giant planets such as Jupiter and Saturn, in the same way as any planetary atmosphere is a very different environment from a comet or the ISM. 

It is also important to quantify what constitutes a \textit{significant} source of a given molecule while evaluating their abiotic versus biotic production mechanisms. We take CH$_3$OH, methanol, as an example. In their Table S2, \cite{seager_prospects_2025} state that there are no known significant abiotic sources of CH$_3$OH on terrestrial planets. However, abiotic mechanisms contribute a non-negligible ($>10\%$) fraction of the total production rate of CH$_3$OH on Earth \citep{Khan2014Methanol, Bates2021Methanol}. At a minimum, this is 20 times higher as a fractional contribution than existing experiments for DMS in a gas mixture similar to the atmosphere of K2-18~b. Thus, the classification of DMS versus CH$_3$OH under the criterion of \textit{No Known Significant False Positives (Context)} in \cite{seager_prospects_2025} (Table 1) appears inconsistent. Overall, it is important that any standard of assessment adopted for biosignatures is applied uniformly across all candidate biosignature molecules.

\section{Methods}
\label{sec:methods}

In this work we perform a wide array of atmospheric retrievals on the JWST transmission spectrum of K2-18~b in the 0.8-12~$\mu$m range using a number of atmospheric retrieval frameworks. This sub-Neptune  \citep{Montet2015, Cloutier2019_K218b_mass} has a mass of 8.63 $\pm$ 1.35 M$_\oplus$ and a radius of 2.61 $\pm$ 0.09 R$_\oplus$ \citep{Cloutier2019_K218b_mass, benneke_water_2019}. Its transit was observed three times with JWST, once with each of NIRISS SOSS, NIRSpec G395H, and MIRI LRS. These observations led to $\gtrsim 3\sigma$ inferences of CO$_2$, CH$_4$ \citep{madhusudhan_carbon-bearing_2023}, and DMS and/or DMDS \citepalias{madhusudhan_new_2025}. The inferences of CO$_2$ and CH$_4$ and non-detections of other species such as NH$_3$, H$_2$O and CO led to its classification as a possible hycean world \citep{madhusudhan_habitability_2021}. 

The inference of DMS and/or DMDS, potential biosignatures \citep{Domagal-Goldman_2011, seager_biosignature_2013, catling_exoplanet_2018, schwieterman_exoplanet_2018, Tsai_2024_Sulfur}, led to the suggestion of possible biological activity on the planet \citepalias{madhusudhan_new_2025}. However, as discussed in \citetalias{madhusudhan_new_2025}, it is important to consider if there are other species that may contribute to the features attributed to DMS and/or DMDS. This possibility was explored by \citetalias{welbanks_challenges_2026}, who, having considered 90 hydrocarbons, found that three of them resulted in Bayes factors $\ln B \geq 2.5$ in both the \citetalias{madhusudhan_new_2025} MIRI \texttt{JExoRES} and \texttt{JexoPipe} data, comparable to that obtained for DMS and/or DMDS. However, the exploration of \citetalias{welbanks_challenges_2026} focused only on the MIRI observations. Therefore, the robustness of their findings was not assessed against the near-infrared data reported in \citet{madhusudhan_carbon-bearing_2023}. 

In the present work, we conduct a very broad search for chemical species in K2-18~b. Furthermore, our search is agnostic, in the sense that we make no prior assumptions as to the theoretical plausibility of any specific chemical species being present in the atmosphere of K2-18~b. Indeed, we consider a list of 661 species, including nearly all those present in the HITRAN cross-section database, several of which are unlikely to be present in the atmosphere of K2-18~b based on present understanding of chemical and physical processes. We then investigate any model preference for them across all the data available. We first systematically investigate whether the MIRI \texttt{JExoRES} and \texttt{JexoPipe} datasets provide support for their presence in the atmosphere of K2-18~b. For those for which at least moderate model preference is consistently found, we then verify whether this is supported also by the near-infrared data presented in \citet{madhusudhan_carbon-bearing_2023}, considered independently as a separate follow-up step. We consider as promising candidates the species which are highest-ranked under our workflow, reaching noteworthy preference across all datasets considered independently. However, we note that in this initial work we do not account for the effect of the multiplicity problem, as we discuss further in Section \ref{sec:discussion}. Previous retrievals on these datasets in \citetalias{madhusudhan_new_2025} and \citetalias{welbanks_challenges_2026} were carried out with the \texttt{AURA} retrieval code and its variant \texttt{Aurora}. For robustness in the present work, we carry out the majority of our retrievals with the retrieval code \texttt{POSEIDON} \citep{macdonald_hd_2017, macdonald_poseidon_2023}. However, in order to ensure our results are independent of the retrieval framework used, we perform multiple cross-checks of promising species with two other independent retrieval codes, \texttt{petitRADTRANS} (\texttt{pRT}, \citealp{molliere_petitradtrans_2019, nasedkin_atmospheric_2024}) and \texttt{VIRA} \citep{constantinou_VIRA_2024}.  
 
\subsection{Observations}

We consider JWST observations of K2-18~b  in total spanning a wavelength range between $\sim$0.8-12 \textmu m taken as part of JWST GO Program 2722 (PI: N. Madhusudhan). For the present analysis, we first consider observations between $\sim$5-12 \textmu m taken with JWST MIRI \citep{Kendrew2015, Bouwman2023} and presented by \citetalias{madhusudhan_new_2025}. The observations took place on April 25-26 2024 over 5.85 hours, with the primary transit itself taking 2.68 hours. For the present work we consider both the \texttt{JExoRES} and \texttt{JexoPipe} reductions of the above MIRI observation. \citetalias{madhusudhan_new_2025} considered different binning prescriptions and found the spectrum to be consistent between them. In this work, we use the nominal binning case of \citetalias{madhusudhan_new_2025}, where the native-resolution spectrum was binned to a width of 0.2 \textmu m or 5 pixels, whichever contained the most pixels. This prescription resulted in reduced $\chi^2$ values of 1.06 and 1.16 with a flat line fit for \texttt{JExoRES} and \texttt{JexoPipe}, respectively. We note that the value of the reduced $\chi^2$ is known to depend on the binning prescriptions adopted, and does not take into account the wavelength distribution of residuals. For these and other reasons, its usage to assess model fits to exoplanet spectra is generally discouraged \citep[e.g][]{welbanks_application_2023}. 

We also consider observations with NIRISS \citep{Doyon2012}, spanning a wavelength range of $\sim$0.8-2.8 \textmu m, combined with NIRSpec G395H \citep{Ferruit2012, Birkmann2014} observations encompassing the $\sim$3-5 \textmu m range, as presented by \citet{madhusudhan_carbon-bearing_2023}. The NIRSpec G395H observations were taken on January 20-21 2023 with a total observing time of 5.3 hours, while the NIRISS observations were made on June 1 2023 over a span of 4.9 hours, each observing one primary transit of K2-18~b. The resulting spectra, as presented in \citetalias{madhusudhan_carbon-bearing_2023}, were binned at 2 pixels per bin for NIRISS, while being kept at the native (1 pixel) resolution for NIRSpec.

\subsection{Retrieval Set-up}

We carry out a retrieval analysis of the above data following the approach of \citetalias{madhusudhan_new_2025}, in order to investigate the model preferences for a wide array of molecules. We specifically consider three independent retrieval frameworks: \texttt{VIRA} \citep{constantinou_VIRA_2024}, \texttt{petitRADTRANS} \citep{molliere_petitradtrans_2019, nasedkin_atmospheric_2024}, and \texttt{POSEIDON} \citep{macdonald_hd_2017, macdonald_poseidon_2023}. We note that \texttt{VIRA} is the latest development of the same family of retrieval codes as the \texttt{Aurora} retrievals used by \citetalias{welbanks_challenges_2026} and the \texttt{AURA} retrievals used in \citetalias{madhusudhan_new_2025}. On the other hand, \texttt{POSEIDON} was developed independently and does not share the same architecture as the \texttt{AURA} family of retrieval codes. All three retrieval codes model the terminator atmosphere as a 1D column in hydrostatic equilibrium with uniform chemical abundances and a parametric $P$-$T$ profile \citep{madhusudhan_temperature_2009}. Each retrieval code uses \texttt{pyMultiNest} \citep{feroz_multinest_2009, buchner_x-ray_2014} for parameter estimation. We have found this to be robust for the dimensionality of the retrievals in this work across different choices for numbers of live points and sampling efficiencies. 
We compute our forward models at average resolving powers of R $\sim 15\,000$ for the NIR and R $\sim 8,000$ for the MIR. Our choice is informed by previous work \citep[e.g.,][]{felix_competing_2025} showing that using higher resolution models (e.g., \citealp{beatty_sulfur_2024, bello-arufe_methane_2025, ahrer_escaping_2025}) does not have a significant impact on retrieval results.  We give our forward model specifications, as well as the settings adopted for other retrieval parameters, in Table \ref{tab:other_settings} in Appendix \ref{sec:priors}.
The mixing ratios of molecular species beyond H$_2$ and He are treated as independent free parameters, with log-uniform priors. The schematic for our framework is shown in Figure \ref{Flow chart figure}.

\subsubsection{Species selection}
\label{sec:species_selection}
We carry out a wide-ranging and agnostic analysis of the MIRI (\citetalias{madhusudhan_new_2025}) and NIRISS/NIRSpec \citep{madhusudhan_carbon-bearing_2023} data for K2-18~b. As shown in Figure \ref{Flow chart figure}, we first conduct a search for 661 species in the MIRI data, using an extension of \texttt{POSEIDON} enabling sequential retrievals. We run retrievals on both the \texttt{JExoRES} and \texttt{JexoPipe} datasets. We then identify the species for which a log-Bayes factor $\ln B \geq 2.0$ is obtained from both the \texttt{JExoRES} and the \texttt{JexoPipe} MIRI data. 
We note that the standard for moderate evidence in \citet{trotta_bayes_2008} is indicated as $\ln B \geq 2.5$. However, this does not consider the presence of uncertainties in the estimation of $\ln B$, due to both random and systematic effects, e.g., due to the use of different retrieval codes. We estimate that this uncertainty can result in variations up to $\ln B \sim 0.5$. We note that this value corresponds to average range spanned by the $\ln B$ values obtained for any given species and dataset through different retrieval codes, reported in Tables \ref{tab:checkMIRIbig12} and \ref{tab:NIR_results}. As our aim is to identify all candidate species which may be able to explain the excess absorption in the spectrum of K2-18~b, in order to be conservative, we pose our threshold for potential moderate evidence at $\ln B = 2.0$.
For the set of species crossing this threshold with \texttt{POSEIDON}, we then verify whether $\ln B \geq 2.0$ model preference is also obtained in retrievals with \texttt{petitRADTRANS} and \texttt{VIRA}. We thus obtain a restricted set of candidate species which reach the $\ln B \geq 2.0$ threshold across all pipelines and retrieval codes considered. For each of these, we then carry out similar retrievals with each of \texttt{VIRA}, \texttt{petitRADTRANS}, and \texttt{POSEIDON} on the near-infrared dataset at native resolution, and again we identify the species which appear promising in the NIR dataset. 

\subsubsection{Model specification}

In order to explore model preferences for the 661 species, we build a canonical model for each. Following the approach developed in \citetalias{madhusudhan_new_2025}, and later replicated in \citetalias{welbanks_challenges_2026}, we define the canonical model for chemical constituent X as containing three chemical constituents: CH$_4$, CO$_2$, and X, where X is any species or combination of species for which the model preference is to be derived. In \citetalias{madhusudhan_new_2025}, canonical models were constructed where X was considered to be DMS, or DMDS, or the combination of DMS and DMDS. In \citetalias{welbanks_challenges_2026}, the set of explored species was expanded, letting X be any of 90 molecules considered. In the present study, X is any of the 661 species considered. The model preference values reported are with respect to the baseline model only including CO$_2$ and CH$_4$. We quote model preference as the natural logarithm of the Bayes factor ($\ln B$).

The 661 species, listed in Appendix \ref{sec:longtable}, include all those present in the opacity database originally built into \texttt{POSEIDON} and available from its repository, as well as most of the HITRAN \citep{HITRAN_2022_Database} cross-section database, including species from all categories listed therein. When available, we adopt the cross-sections at $T \approx 300$ K and $P \approx 1$ bar; otherwise, we use cross-sections for the closest available temperature. We excluded species for which no opacity data was available in the mid infrared, or, as in the case of formaldehyde, it was available only for very high temperatures ($T =998$K). Furthermore, 3 hydrofluoroethers whose molecular weight we were not able to independently verify were also excluded. We thus obtained our final set of  661 species. For wavelength regions not covered by the available cross-section data, we assume no opacity from the relevant absorber. The pressure, temperature and wavelength range for each cross-section obtained from HITRAN is reported in Table \ref{tab:all-molecules} for those species resulting in noteworthy model preferences.

For all our retrievals, we adopt the \cite{madhusudhan_temperature_2009} pressure-temperature profile, like \citetalias{madhusudhan_new_2025} and \citetalias{welbanks_challenges_2026} do, which introduces 6 free parameters to our retrievals. We also allow for patchy clouds and hazes, following the \citet{macdonald_hd_2017} and \citet{pinhas_h2o_2019} parametrisation. This includes 4 free parameters: cloud-top pressure $P_{\rm c}$ for the cloud deck; Rayleigh enhancement factor $a$ and wavelength dependence $\gamma$ (such that the scattering cross-section $\sigma$ depends on wavelength $\lambda$ as $\sigma \sim a \lambda^\gamma$) for the hazes, with $a=1$ and $\gamma = -4$ corresponding to Rayleigh scattering; and a fractional cloud and haze coverage parameter $\phi$, with $\phi =0$ indicating a fully clear atmosphere, and $\phi = 1$ a uniformly cloudy/hazy one. Finally, we include a free reference pressure $P_{\rm ref}$, i.e., the pressure to which the white-light radius ($R_{\rm p} = 2.61 R_\oplus$) corresponds. This leads to 11 free parameters, plus the 3 molecular mixing ratios, for a total of 14 free parameters in most canonical model \texttt{POSEIDON} retrievals. This grows to 15 free parameters when considering the near-infrared data and allowing for an offset between the NIRSpec and the NIRISS datasets. We show our model specifications and retrieval settings in Table \ref{tab:other_settings}, and our priors in Table \ref{tab:all_priors}, both in Appendix \ref{sec:priors}. 
As \texttt{petitRADTRANS} works with mass mixing ratios rather than volume ones, we choose a prior $\mathcal{U}(-11, -0.1)$ to better reflect, in a low-mean molecular weight atmosphere, the $\mathcal{U}(-12, -0.3)$ prior in volume mixing ratios used in the \texttt{POSEIDON} and \texttt{VIRA} retrievals.

\section{Results}
\label{sec:results}

In this work we performed a systematic search for the spectral signatures of 661 species in the atmosphere of exoplanet K2-18~b, using the three published datasets: the \citetalias{madhusudhan_new_2025} \texttt{JExoRES} and \texttt{JexoPipe} data instances for the MIRI LRS spectrum, and the \citet{madhusudhan_carbon-bearing_2023} NIRISS-NIRSpec spectrum. We first explored which molecules the MIRI spectrum may be showing moderate (or stronger) evidence for, and then investigated if such evidence may also be present in the near-infrared (NIR) range. In this section, we describe the outcome of this process. We start by considering the results obtained from the MIRI \texttt{JExoRES} dataset. We then compare these results with those obtained from the MIRI \texttt{JexoPipe} dataset, and as the final step in our workflow we search for the most promising species in the NIR dataset.

\subsection{Canonical retrievals with MIRI 
\label{sec:jres}
\texttt{JExoRES} data}
We begin by considering the MIRI \texttt{JExoRES} data. We first ensure consistency of our framework with established results by reproducing the key findings from \citetalias{madhusudhan_new_2025} and \citetalias{welbanks_challenges_2026}. We then expand our search to a list of 661 species.  

\subsubsection{Reproduction of previous work}
\label{sec:repr_previous}
\citetalias{madhusudhan_new_2025} reported $\sim$3$\sigma$ evidence, corresponding to $\ln B =2.85$-$4.22$, for DMS and/or DMDS, when using the canonical model approach, i.e., comparing a model only including CH$_4$ and CO$_2$ with one also including either DMS, or DMDS, or both at the same time. This was reproduced by \citetalias{welbanks_challenges_2026}. We note that \citetalias{madhusudhan_new_2025} used the \texttt{AURA} retrieval framework \citep{pinhas_retrieval_2018}, and \citetalias{welbanks_challenges_2026} used \texttt{Aurora} \citep{welbanks_aurora_2021, nixon_methods_2024}. Both codes belong to the \texttt{AURA} family, sharing many of the core components, and were developed in the same research group, so the agreement is expected. 
Here, we reproduce the \citetalias{madhusudhan_new_2025} results for the model preferences for DMS and/or DMDS with 3 independent codes, \texttt{POSEIDON} \citep{macdonald_hd_2017, macdonald_poseidon_2023},
\texttt{petitRADTRANS} \citep{molliere_petitradtrans_2019, nasedkin_atmospheric_2024}, and \texttt{VIRA} (\citealp{constantinou_VIRA_2024}), the latter of which also belongs to the \texttt{AURA} retrieval codes family, being its latest version. 
As shown in Table \ref{tab:checkMIRIbig12}, generally consistent results are obtained across the 3 retrieval codes considered in this study, which are in turn consistent with the results reported by \citetalias{madhusudhan_new_2025} and \citetalias{welbanks_challenges_2026}. 
We also consider the three additional species that, according to \citetalias{welbanks_challenges_2026}, result, with both the \texttt{JExoRES} and \texttt{JexoPipe} data, in Bayes factors $\ln B\geq 2.5$,  which, according to Jeffreys' scale (see, e.g., \citealp{trotta_bayes_2008}), corresponds to the minimum threshold for moderate evidence. For these three species as well, we perform consistency checks with the three above-mentioned codes, also finding overall good agreement with the \citetalias{welbanks_challenges_2026} results.

\begin{table*}[]
\centering
\begin{tabular}{l|llllll}
Species           & \multicolumn{2}{c}{\texttt{VIRA}} & \multicolumn{2}{c}{\texttt{pRT}}    &  \multicolumn{2}{c}{\texttt{POSEIDON}}   \\
                  & \texttt{JExoRES}              & \texttt{JexoPipe}             & \texttt{JExoRES}         & \texttt{JexoPipe}        & \texttt{JExoRES}         & \texttt{JexoPipe}        \\ \hline
Chloroethane      & $3.5$ & $4.4$ & $3.3$ & $4.2$ & $3.1$  & $4.0$  \\
Propyne           & $3.1$ & $3.1$ & $2.3$ & $2.7$ & $2.6$  & $2.7$ \\
Dichloromethane   & $2.7$ & $2.9$ & $2.0$  & $2.6$ & $2.6$  & $2.8$ \\
Methacrylonitrile & $2.7$ & $2.6$ & $2.6$ & $2.7$ & $2.6$   & $2.4$ \\
Bromoethane       & $2.9$ & $2.2$ &  $2.5$ & $2.0$ & $2.6$ & $2.0$ \\ 
Cyclohexane       & $2.7$ & $3.0$ & $1.6$  & $2.1$ & $2.5$  & $3.0$ \\
Butane            & $2.6$ & $2.4$ & $2.4$  & $2.4$ & $2.5$  & $2.1$  \\
Cyclopentane      & $2.3$ & $2.1$ & $2.0$ & $2.1$ & $2.4$  & $2.2$  \\
DMS               & $2.6$ & $2.8$ & $2.0$ & $2.5$ & $2.2$  & $2.5$ \\
Allyl chloride    & $2.3$ & $2.6$ & $1.9$ & $2.0$ & $2.1$ & $2.1$  \\
DMDS              & $3.5$ & $3.0$ & $1.8$ & $1.9$ & $2.3$ & $1.9$   \\ \hline
\end{tabular}

    \caption{Values of $\ln B$ against the baseline CH$_4$+CO$_2$ model from \texttt{petitRADTRANS}, \texttt{VIRA} and \texttt{POSEIDON} for the 10 molecules which reach the $\ln B = 2.0$ threshold in the \texttt{POSEIDON} retrievals using both \texttt{JExoRES} and \texttt{JexoPipe} MIRI data, as well as for DMDS, found at $\ln B \geq 2.5$ in both MIRI datasets by \citetalias{madhusudhan_new_2025}. Typical uncertainties on $\ln B$ are $\sim \pm 0.5$.  As a guide, the \citet{sellke_calibration_2001} and \citet{trotta_bayes_2008} conversion yields $\ln B = 1.0 \iff 2.0\sigma$, $\ln B = 2.0 \iff 2.5\sigma$ and $\ln B = 3.0 \iff 3.0\sigma$.}
\label{tab:checkMIRIbig12}
\end{table*}

\subsubsection{Exploration of additional species}
We conduct a broad exploration of additional chemical species in search of those that may explain the features in the MIRI LRS data besides DMS and/or DMDS, reported in \citetalias{madhusudhan_new_2025}. \citetalias{welbanks_challenges_2026} considered 90 species in addition to DMS and DMDS, focusing on hydrocarbons. We now significantly expand upon the \citetalias{madhusudhan_new_2025} and \citetalias{welbanks_challenges_2026}  work, considering nearly all species for which cross-sections are available in the HITRAN database. In particular, we consider 556 species across the following classes: hydrocarbons; nitriles, amines and other nitrogenated hydrocarbons; sulfur-containing species; alcohols, ethers and other oxygenated hydrocarbons; bromocarbons, hydrobromocarbons and halons; chlorocarbons and hydrochlorocarbons; chlorofluorocarbons (CFCs); fully fluorinated species; halogenated alcohols and ethers; hydrochlorofluorocarbons (HCFCs); hydrofluorocarbons (HFCs); and iodocarbons and hydroiodocarbons. We also include an additional 29 species belonging to other classes, as well as a further 76 originally considered in \texttt{POSEIDON}. Overall, our final database includes a total of 661 species. 

\begin{figure*}
    \centering

        \includegraphics[width=\linewidth]{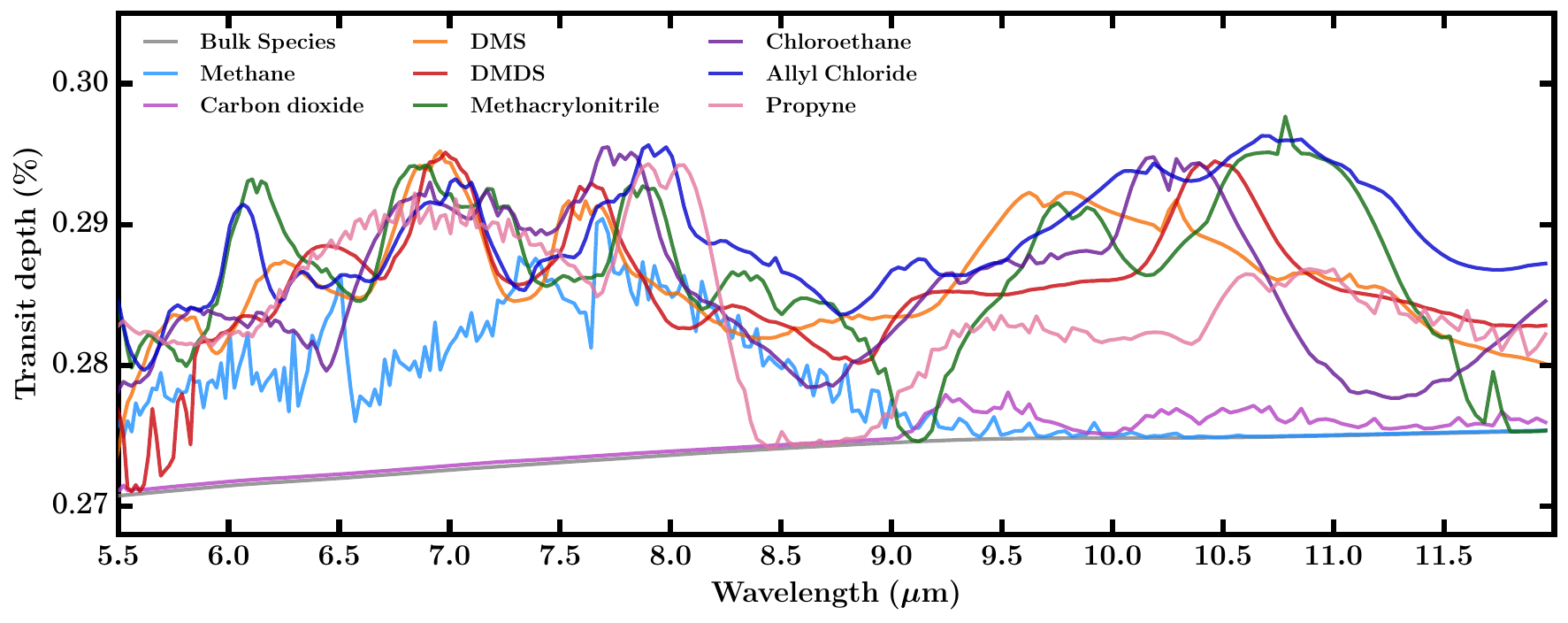}

    \caption{Contributions of molecules in the JWST MIRI range for the transmission spectrum of K2-18~b, assuming 1\% CO$_2$ and CH$_4$, and $10^{-4}$ by volume for each of DMS, DMDS, methacrylonitrile, chloroethane, allyl chloride and propyne.}
    \label{fig:contributions}
\end{figure*}

The model preference values for the species resulting in $\ln B \geq 1.15$ are reported in Table \ref{tab:all-molecules} in Appendix \ref{sec:longtable}, and all remaining species are listed in Appendix \ref{sec:remaining}. We find that 15 of these are preferred by the \texttt{JExoRES} data at $\ln B \geq 2.0$ over a model only including CH$_4$ and CO$_2$.
As mentioned above, our choice of $\ln B  = 2.0$ as a threshold is due to this corresponding to the standard for moderate evidence in Jeffreys' scale, when accounting for a typical uncertainty of up to $\Delta \ln B = \pm 0.5$ in the estimation of the evidence. 
We note, however, that not all of these 15 molecules are known to have abiotic sources on Earth or beyond, with several being primarily biogenic or anthropogenic. We show a contribution plot in the mid-infrared range including some of these species in Figure \ref{fig:contributions}.

Among these 15 molecules, most result in Bayes factors comparable to DMS, spanning a range in $\ln B$ of 1. 
However, it is possible that some of these species may only be preferred by the data from the \texttt{JExoRES} pipeline, and not by the \texttt{JexoPipe} one. Hence, we next proceed to perform the same retrievals on the \texttt{JexoPipe} data, in order to ascertain which species could be reliable substitutes for DMS and/or DMDS to explain the observed mid-infrared features.

\subsection{Canonical retrievals with MIRI \texttt{JexoPipe} data}
\label{sec:jpipe}
As mentioned above, in order to reach robust conclusions, it is important to check that potential evidence for any molecule is not an artifact of a specific data reduction pipeline.

\begin{figure*}[ht]
    \centering
    \includegraphics[width=\linewidth]{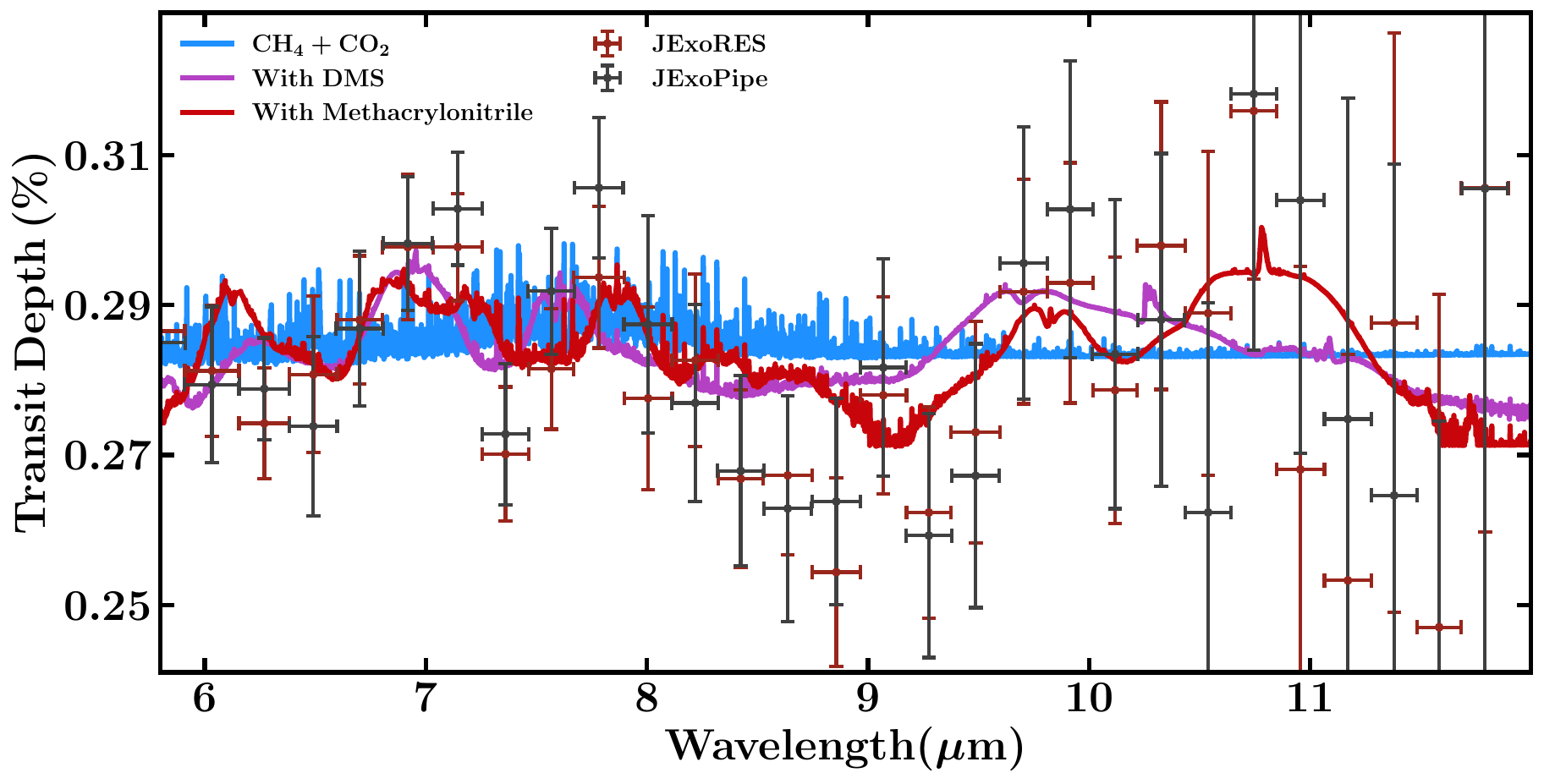}
    \caption{Retrieved spectra with the JWST MIRI \texttt{JexoPipe} data. The solid curves show  median retrieved spectra obtained from the \texttt{POSEIDON} retrievals on MIRI \texttt{JexoPipe} data for the CH$_4$+CO$_2$ baseline, and the canonical models for each of dimethyl sulfide and methacrylonitrile, the two species which are most preferred against the baseline model, when considering all cases described in this work. 
    Also shown are the \texttt{JExoRES} and \texttt{JexoPipe} MIRI observations.} 
    \label{fig:spectral_fits}
\end{figure*}

In order to achieve this additional level of robustness, here we compute the model preferences for each of 661 species we considered for \texttt{JExoRES} with the \texttt{JexoPipe} data as well. For species resulting in $\ln B \geq 1.15$, we report the results in Table \ref{tab:all-molecules} in Appendix \ref{sec:longtable}. We find that, as for \texttt{JExoRES}, 15 species result in log-Bayes factors $\geq 2.0$ over CH$_4$ and CO$_2$ only. Five of these had been found at $\ln B < 2.0$ in the \texttt{JExoRES} dataset: tetramethylsilane (at $\ln B = 1.6$ in \texttt{JExoRES}),  dimethyl carbonate ($\ln B = 1.5$), ethanethiol ($\ln B = 1.4$), 1-chloropentane ($\ln B = 1.2$), and dichloromethylphosphine ($\ln B = 1.0$).
On the other hand, of those for which $\ln B \geq 2.0$ was obtained with the \texttt{JExoRES} dataset, the five that were not found with the \texttt{JexoPipe} spectrum are: dimethyl disulfide (DMDS, $\ln B = 1.9$ with \texttt{JexoPipe}), methacryloyl chloride  and methacrolein (both at $\ln B = 1.3$ with \texttt{JexoPipe}), trans-2-pentene ($\ln B = 1.1$), and 2-butene ($\ln B = 0.8$). We show the median retrieved spectra for DMS and methacrylonitrile in Figure \ref{fig:spectral_fits}.

\begin{figure*}
    \centering
    \includegraphics[width=\linewidth]{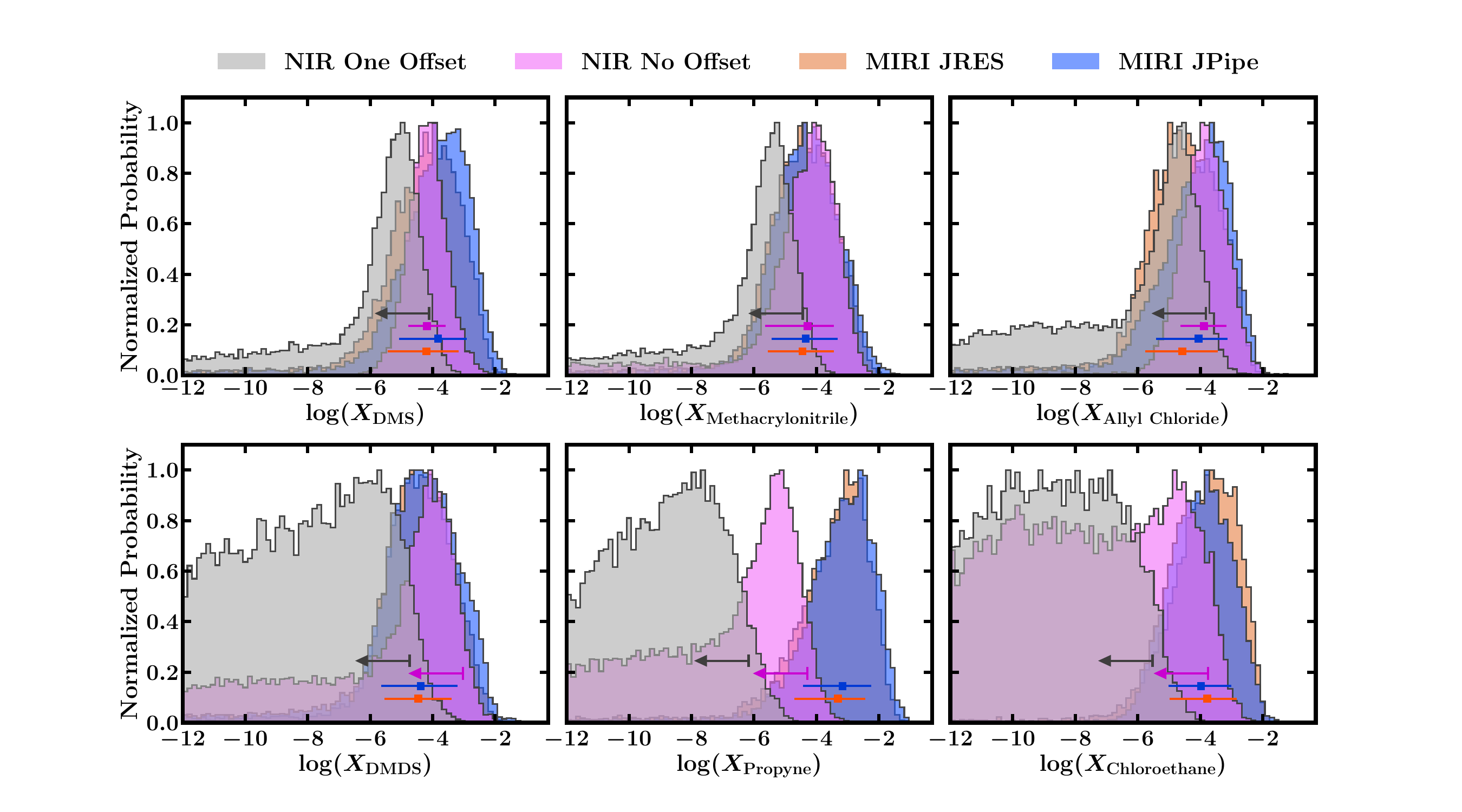}
    \caption{Posterior distributions for the volume mixing ratios of selected species, as obtained from the respective canonical model \texttt{POSEIDON} retrievals on the MIRI \texttt{JExoRES}, MIRI \texttt{JexoPipe}, and NIRISS-NIRSpec spectra of K2-18~b.}
    \label{fig:posteriors}
\end{figure*}

Overall, ten species reach the $\ln B = 2.0$ threshold in both datasets. These are: chloroethane, bromoethane, propyne,  methacrylonitrile, dichloromethane, cyclohexane, butane, cyclopentane, DMS, and allyl chloride. For these ten species, we perform a robustness check with the two additional retrieval codes \texttt{petitRADTRANS} and \texttt{VIRA}, for both \texttt{JExoRES} and \texttt{JexoPipe} data, shown in Table \ref{tab:checkMIRIbig12}. We find that eight species are confirmed to have preference $\ln B \geq 2.0$ in all sets of retrievals, the exceptions being allyl chloride and cyclohexane. 
This leads to a final set of eight molecules for which at least moderate evidence may consistently be inferred in the mid-infrared. The abundances found through \texttt{VIRA} and \texttt{POSEIDON} for these species are reported in Table \ref{tab:abundances}, and the \texttt{POSEIDON} posterior distributions for a subset of these are shown in Figure \ref{fig:posteriors}.

\begin{table*}[t]
\centering
\begin{tabular}{l|ccc|ccc}

\multicolumn{1}{c|}{Species} & \multicolumn{3}{c|}{No Offset} & \multicolumn{3}{c}{One Offset} \\ 

                            & \texttt{VIRA} & \texttt{pRT}  & \texttt{POSEIDON}  & \texttt{VIRA} & \texttt{pRT}  &  \texttt{POSEIDON} \\ \hline
DMS                          & $2.9$         & $2.8$     & $3.0$      &  $0.5$            &  $0.4$  & $0.8$        \\
Methacrylonitrile            & $2.5$        & $1.9$      & $2.0$     &  $0.8$         &  $1.0$  & $0.8$    \\
Allyl Chloride               & $2.2$        & $1.4$     & $1.8$    &   $0.4$      &  $0.2$   & $0.6$    \\
Butane                       & $1.0$       & $0.5$    & $0.9$    &  $-0.3$    &  $-0.6$  & $-0.2$  \\
Bromoethane                 &  $0.8$       & $0.6$     &  $0.3$   &  $-0.1$    & $-0.3$    & $0.0$  \\
Cyclopentane                 & $0.6$       & $0.6$     & $0.2$    &     $-0.3$   &    $0.0$ & $-0.4$  \\
Cyclohexane                  & $0.4$       & $0.6$    & $0.4$    &  $-0.4$    &  $-0.1$   & $-0.5$   \\ 
Chloroethane                 & $0.2$       & $-0.1$     & $-0.1$    &   $-0.4$   &    $-0.2$    & $-0.1$   \\ 
Propyne                      & $0.1$       & $0.9$      & $0.7$       &   $-0.4$     &   $-0.2$   & $-0.4$   \\
Dichloromethane              & $0.0$      & $-0.4$      & $-0.1$       &      $0.0$  &   $-0.1$   & $-0.4$  \\
DMDS                         & $0.7$       & $1.0$       & $0.8$     &  $-0.3$     &   $-0.1$  & $0.0$   \\
\end{tabular}
\caption{Bayes factors $\ln B$ in the near-infrared for the 11 species reported in Table \ref{tab:checkMIRIbig12}, relative to a CH$_4$+CO$_2$ model. Typical uncertainties on $\ln B$ are $\sim \pm 0.5$. As a guide, the \citet{sellke_calibration_2001} and \citet{trotta_bayes_2008} conversion yields $\ln B = 1.0 \iff 2.0\sigma$, $\ln B = 2.0 \iff 2.5\sigma$ and $\ln B = 3.0 \iff 3.0\sigma$.}
\label{tab:NIR_results}
\end{table*}

\begin{table*}[]
\centering
\begin{tabular}{l|llll|llll}
\multicolumn{1}{c|}{Species} & \multicolumn{4}{c|}{\texttt{VIRA}} & \multicolumn{4}{c}{\texttt{POSEIDON}} \\
& \multicolumn{2}{c}{MIR}                              & \multicolumn{2}{c|}{NIR} &  \multicolumn{2}{c}{MIR}                              & \multicolumn{2}{c}{NIR} \\
                             & \multicolumn{1}{c}{\texttt{JRES}} & \multicolumn{1}{c}{\texttt{JPipe}} & \multicolumn{1}{c}{No Offset} &\multicolumn{1}{c|}{1 Offset} & \multicolumn{1}{c}{\texttt{JRES}} & \multicolumn{1}{c}{\texttt{JPipe}} & \multicolumn{1}{c}{No Offset} &\multicolumn{1}{c}{1 Offset}                        \\ \hline
DMS               & $-3.66^{+1.10}_{-1.42}$ & $-3.65^{+1.12}_{-1.33}$ & $-4.42^{+0.59}_{-0.61}$ & $<-4.15$ & $-4.20^{+1.03}_{-1.24}$ & $-3.83^{+0.92}_{-1.25}$ & $-4.18^{+0.60}_{-0.60}$    & $ <-4.12$\\
Methacrylonitrile & $-4.13^{+1.26}_{-1.24}$ & $-4.04^{+1.28}_{-1.30}$ & $-4.60^{+0.72}_{-0.70}$                & $<-4.24$ & $-4.44^{+1.01}_{-1.11}$ & $-4.34^{+1.03}_{-1.09}$ & $-4.28^{+0.83}_{-1.35}$    & $ <-4.44$\\
Allyl chloride    & $-3.97^{+1.18}_{-1.48}$ & $-3.98^{+1.19}_{-1.37}$ & $-3.96^{+0.68}_{-0.69}$               & $<-3.63$ & $-4.58^{+1.13}_{-1.18}$ & $-4.06^{+0.94}_{-1.36}$ & $<-2.75$                    & $<-3.82$\\ 
Butane            & $-3.68^{+1.21}_{-1.41}$ & $-3.37^{+1.11}_{-1.46}$ & $< -3.30$               & $<-5.17$ & $-4.02^{+1.02}_{-1.32}$ & $-3.92^{+1.08}_{-1.26}$ & $ <-3.30$                   & $ <-5.42$\\
Bromoethane       & $-3.56^{+1.03}_{-1.41}$ & $-3.64^{+1.15}_{-1.52}$ &  $<-2.86$     &    $<-4.72$ & $-3.92^{+0.97}_{-1.24}$   &$-3.89^{+1.11}_{-1.60}$  & $<-2.92$                    & $<-5.13$ \\
Cyclopentane      & $-3.67^{+1.32}_{-1.35}$ & $-3.32^{+1.27}_{-1.48}$ & $<-2.88$      & $<-5.16$ & $-3.91^{+1.10}_{-1.09}$ & $-3.55^{+1.10}_{-1.33}$ & $<-3.17$                    & $<-5.44$\\
Cyclohexane       & $-2.28^{+0.84}_{-1.50}$ & $-2.11^{+0.73}_{-1.37}$ & $<-3.70$                & $<-5.73$ & $-2.93^{+1.02}_{-1.25}$ & $-2.39^{+0.78}_{-1.42}$ & $<-3.52$                   & $<-5.86$\\
Chloroethane      & $-3.38^{+1.03}_{-1.29}$ & $-3.43^{+1.03}_{-1.36}$  & $< -3.42$              & $<-5.06$ & $-3.78^{+0.94}_{-1.21}$ & $-3.97^{+0.97}_{-1.04}$ & $<-3.73$                   & $<-5.53$\\
Propyne           & $-2.84^{+0.97}_{-1.29}$ & $-2.91^{+1.03}_{-1.36} $ & $<-4.14$               & $<-6.02$ & $-3.31^{+0.89}_{-1.39}$ & $-3.16^{+0.92}_{-1.26}$ & $<-4.29$                   & $<-6.18$\\
Dichloromethane   & $-2.60^{+0.73}_{-1.09}$ & $-2.81^{+0.77}_{-1.08}$ & $<-2.87$                & $<-4.26$ & $-2.71^{+0.69}_{-1.20}$ & $-2.96^{+0.78}_{-1.12}$ & $<-3.15$                   & $<-4.36$\\
DMDS              & $-3.35^{+1.00}_{-1.25}$ & $-3.40^{+1.10}_{-1.32}$  & $<-3.26$               & $<-4.82$ & $-4.46 ^{+1.06}_{-1.09}$ & $-4.45^{+1.10}_{-1.13}$ & $<-3.04$                  & $<-4.74$\\

\end{tabular}
\caption{Abundance estimates for the 11 species listed in Table \ref{tab:checkMIRIbig12} obtained with \texttt{VIRA} and \texttt{POSEIDON} canonical model retrievals on the three datasets considered in this work. When a dataset provides $\ln B \geq 2.0$ support for a species with the relevant retrieval code, the median retrieved abundance and $1\sigma$ error bars are reported; otherwise, the $2 \sigma$ upper limit is indicated.} 

\label{tab:abundances}
\end{table*}

\subsection{Canonical retrievals with NIR data}
\label{sec:NIR}
A near-infrared  spectrum is also available for K2-18~b, obtained with the JWST NIRISS SOSS and NIRSpec G395H instruments \citep{madhusudhan_carbon-bearing_2023}. Any claim of evidence for a trace species, such as DMS, should be verified with multiple lines of evidence, e.g. with different observations or different instruments. This has so far only been the case for DMS, for which \citet{madhusudhan_carbon-bearing_2023} find model preferences of up to $2.4 \sigma$ ($\ln B = 1.7$), in their no-offset scenario, but below $2 \sigma$ ($\ln B < 0.9$) in the one-offset case. \citet{hu_water-rich_2025} later confirmed that model preferences for DMS of up to 2.7$\sigma$ ($\ln B = 2.3)$ over a model only including six standard CNO species (CH$_4$, CO$_2$, CO, NH$_3$, HCN, H$_2$O) can be obtained in the NIR, including offsets between all considered detectors. 

For consistency with the approach we adopted with the MIRI observations, we here adopt the same canonical model configuration as in \citetalias{madhusudhan_new_2025}, \citetalias{welbanks_challenges_2026} and our Section \ref{sec:jres} and Section \ref{sec:jpipe}, containing three chemical constituents: CH$_4$, CO$_2$, and X, where X is the species for which the model preference is to be derived. The model preference values we quote will hence be for these canonical models against a base model only including CH$_4$ and CO$_2$. Furthermore, as the motivation to search for DMS in the atmosphere of K2-18~b was originally the result of a potential inference of it in the no-offset scenario, we adopt no offsets at first here too, to verify if such a possibility exists for other species. As simpler models generally lead to increased evidence, the values in this work should be considered upper bounds on the evidence provided by the NIR data for each considered species.

We thus proceed to verifying which of the eight species for which the $\ln B = 2.0$ threshold was consistently cleared in the mid-infrared data are also supported by the near-infrared data. Additionally, we also verify whether the near-infrared data provide support for the two species identified at $\ln B \geq 2.5$ in both MIRI \texttt{JExoRES} and \texttt{JexoPipe} datasets by \citetalias{welbanks_challenges_2026} or \citetalias{madhusudhan_new_2025} for which we do not consistently find  $\ln B \geq 2.0$  when considering the two mid-infrared datasets and three retrieval codes: cyclohexane and DMDS. For completeness, we also include allyl chloride in our NIR analysis, to avoid it being the only species from Table \ref{tab:checkMIRIbig12} excluded from this analysis. This leads to a total of 11 species to be verified against the NIR data. 

We carry out atmospheric retrievals on this set of species using \texttt{VIRA}, \texttt{petitRADTRANS}, and \texttt{POSEIDON}. We show the obtained model preference values in Table \ref{tab:NIR_results}, finding that only one of the molecules is supported at $\ln B \geq 2.0$ in the NIR data across all three codes, DMS  ($\ln B = 2.8$-$3.0$). One more species, methacrylonitrile, clears the $\ln B \geq 2.0$ threshold with \texttt{VIRA} and \texttt{POSEIDON}, but not with \texttt{pRT}, and another, allyl chloride, only does so with \texttt{VIRA}.

DMS, thus, is the only species for which our workflow yields $\ln B \geq 2.0$ across all considered datasets, i.e., two datasets corresponding to independent reductions of the \citetalias{madhusudhan_new_2025} MIRI LRS spectrum and one corresponding to the \citetalias{madhusudhan_carbon-bearing_2023} NIR spectrum, and  with all retrieval frameworks used. 

As discussed above, the significance values found here ought to be considered as upper bounds on the evidence for these species in the NIR, as considering more complex models can lead to lower evidence. For example, as reported in \citet{madhusudhan_carbon-bearing_2023}, when considering their preferred canonical model with one offset, the significance for DMS was $\sim$1$\sigma$ compared to 2.4$\sigma$ in the zero-offset case. Similarly, in our present retrievals, none of the molecules considered above reach a $\ln B = 2.0$ preference when allowing for an offset between the NIRISS and NIRSpec detectors.

\subsection{Maximal Retrievals}
\label{sec:maximal}
Thus far, we have adopted a canonical model approach. This approach offers the significant benefit of allowing a fast and broad exploration of a wide set of chemical species (up to 661, in this work). However, this comes at the expense of using a simple model, which in this case only contains two baseline chemical species (CO$_2$ and CH$_4$) in addition to the extra absorber which is investigated. This risks resulting in degeneracies between different absorbers being overlooked. Furthermore, if the baseline set of chemical species is incomplete, it may result in overestimating the detection significance for the extra absorber.

At the other end of the complexity spectrum are maximal model approaches (e.g., \citetalias{madhusudhan_new_2025}). These include a broader set of chemical species at once (as broad as 20, in \citetalias{madhusudhan_new_2025}), and as such they offer the benefit of exploring degeneracies between different absorbers, as well as the effects of including combinations of species. However, if the set of species considered is exceedingly broad, detection significances found through maximal models will be underestimated. 

For completeness, we report here the results of maximal retrievals on all datasets considered in this study, carried out with \texttt{POSEIDON}. We verify the evidence for any species or set of species by comparing the maximal model to a nested model with the relevant absorber(s) removed, following the standard \citep{benneke_how_2013} approach. For all retrievals, we preserve identical setups to those employed for the corresponding canonical models, the only change being the absorbers included.
\subsubsection{Mid infrared}
We consider a maximal set up including six CNO chemical species which may be expected in a hydrogen-rich temperate atmosphere (CO$_2$, CH$_4$, H$_2$O, CO, NH$_3$ and HCN) as well as the 11 species listed in Table \ref{tab:checkMIRIbig12}. We first verify whether any model preference is present for excess absorption beyond CH$_4$ and CO$_2$, by removing all 15 remaining species. We find weak evidence towards this being the case with the \texttt{JExoRES} data, at $\ln B = 1.9$ (corresponding to 2.5$\sigma$ using the conversion reported in \citealp{sellke_calibration_2001, trotta_bayes_2008, benneke_how_2013}), while moderate evidence ($\ln B = 2.2 $, 2.6$\sigma$) can be reached with \texttt{JexoPipe}. We note that in both cases this is lower than found with the most preferred canonical models, each including one species beyond CH$_4$ and CO$_2$. This is due to the complexity penalty introduced in calculating the Bayesian evidence ($\ln Z$), such that a maximal model including several degenerate species results in a lower evidence than a smaller model (e.g., a canonical one) which provides an equally good fit to the data.

The removal of the 11 species in Table \ref{tab:checkMIRIbig12} from the maximal model does not result in a meaningful change in model evidence when using the \texttt{JExoRes} data from either pipeline, with a Bayes factor in favour of the presence of those 11 species of $\ln B = 0.8$ (1.9$\sigma$).
On the other hand, slightly higher significances are found when considering the \texttt{JexoPipe} data. In this case, we find weak evidence ($\ln B = 1.3$, 2.2$\sigma$) towards the species listed in Table \ref{tab:checkMIRIbig12} being present.

\subsubsection{Near infrared}
We next perform a similar analysis with the near-infrared data. We begin with a no offset case, consistently with the approach we adopted for the canonical model retrievals. In this case, our maximal model includes the six aforementioned CNO species (CO$_2$, CH$_4$, H$_2$O, CO, NH$_3$ and HCN), and the two species which the canonical model retrievals with \texttt{POSEIDON} revealed to be promising (DMS and methacrylonitrile). This maximal model is preferred at $\ln B = 3.2$ (3.0$\sigma$) over one only including the six CNO species. However, using the maximal model as a reference, only weak evidence ($\ln B = 1.3$, 2.2$\sigma$) is found for DMS individually, and none is found for methacrylonitirle, indicating they are largely degenerate with each other. 

\begin{figure*}
    \centering
    \includegraphics[width=\linewidth]{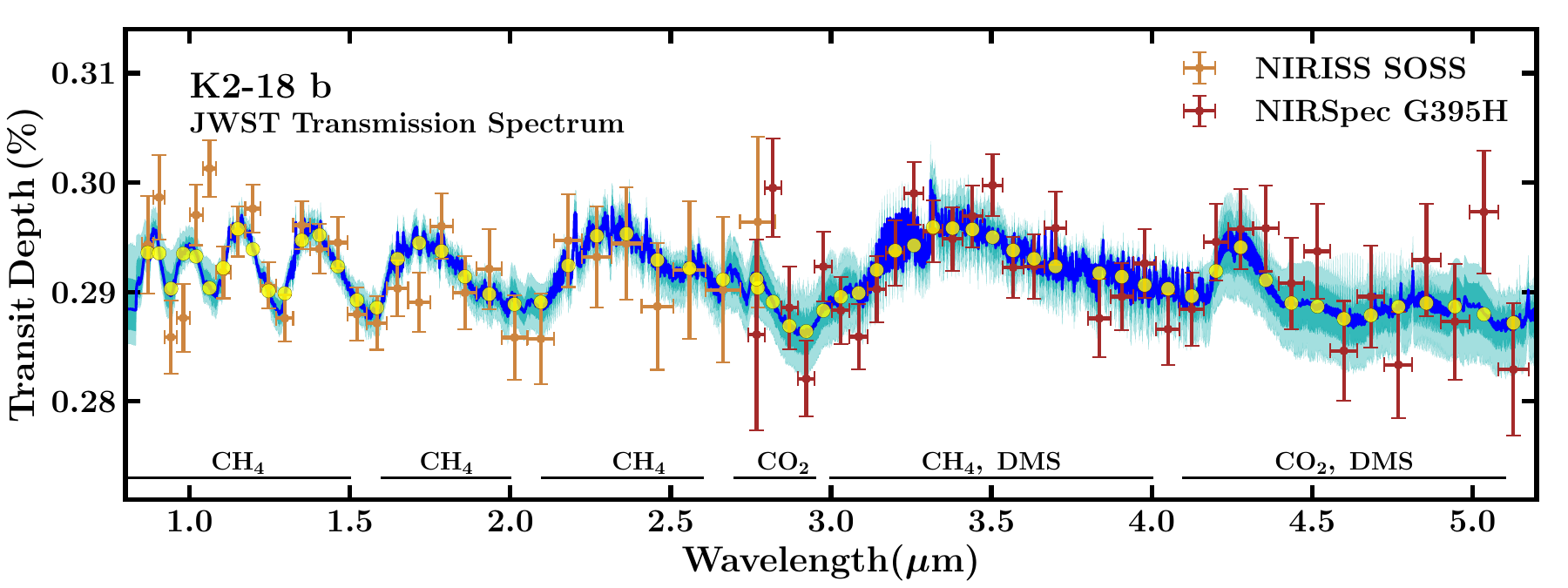}
    \caption{Median retrieved spectrum with \texttt{POSEIDON} (blue line) and 1 and 2$\sigma$ uncertainties (shaded regions) in the near-infrared for a model including six standard CNO species (CO$_2$, CH$_4$, NH$_3$, CO, HCN and H$_2$O) and DMS, with one offset. The contributions of the various absorbers are shown in Figure 5 of \citet{madhusudhan_carbon-bearing_2023} and Figure 3 of \citet{hu_water-rich_2025}}
    \label{fig:nir_spectral_fit}
\end{figure*}

When allowing for an offset, the above maximal model including the six standard CNO species (CO$_2$, CH$_4$, H$_2$O, CO, NH$_3$ and HCN) and  DMS and methacrylonitrile results in weak preference ($\ln B = 1.3$, 2.2$\sigma$) over the model with the standard species only, while no statistically significant preference is found for either of DMS or methacyrlonitrile alone compared to the maximal model. We show the median retrieved spectrum for a one-offset model including the 6 standard CNO species as well as DMS, the species that is overall most preferred when considering our work in its entirety, in Figure \ref{fig:nir_spectral_fit}. 

\subsection{Overall significances}
Overall, we find that, of 661 initial species, 15 crossed the $\ln B = 2.0$ threshold for model preference against each of the MIRI \texttt{JExoRES} and MIRI \texttt{JexoPipe} data, with \texttt{POSEIDON}. Of these, eight crossed the threshold in both mid-infrared datasets with all of \texttt{VIRA}, \texttt{petitRADTRANS} and \texttt{POSEIDON}, and were further investigated in the near-infrared. Among these, only DMS reached the $\ln B = 2.0$ threshold in the near-infrared data with all three retrieval codes used, when considering no offsets between detectors.

Beyond the canonical model retrievals, where the baseline model was only CH$_4$ and CO$_2$, we also considered maximal model retrievals with \texttt{POSEIDON}, which overall confirmed the inferences from the canonical models. In particular, for the MIRI data, we adopted a maximal model with 17 species, including the 6 standard CNO species, i.e.,  CH$_4$, CO$_2$, H$_2$O, CO, NH$_3$, and HCN, as well as the eleven species listed in Table \ref{tab:checkMIRIbig12}. We found that this model is somewhat preferred over the CH$_4$ and CO$_2$ baseline, at $\ln{B} = 1.9$ (2.5$\sigma$) with the \texttt{JExoRES} data and $\ln{B} = 2.2$ (2.6$\sigma$) with \texttt{JexoPipe}. 
Only when using the \texttt{JexoPipe} data, weak evidence ($\ln{B} = 1.3$, 2.2$\sigma$) for the maximal model is also found against one including all 6 standard CNO species. 

When considering the NIR data without offsets, we also find moderate preference for the presence of DMS and/or methacrylotirile. In particular, a maximal model including these two species as well as the six CNO species was preferred at $\ln B = 3.2$ (3.0$\sigma$) over one only including the 6 expected CNO species. However, DMS and methacrylonitrile remained mostly degenerate, with only DMS resulting in some weak evidence when considered individually, using the maximal model as reference. When considering one offset between the NIRSpec and NIRISS instruments, only weak preference is found for DMS and/or methacrylonitrile. Specifically, a model also including these two species is preferred at $\ln B = 1.3$ (2.2$\sigma$) against one only including the six standard CNO species, and the degeneracy between the two species persists.

We highlight that this analysis shows the potential of atmospheric retrievals. When used with appropriate robustness checks and across a few datasets, they allow an agnostic list of 661 candidate molecules to be reduced to just a handful of promising candidates. In this work, requiring Bayes factors compatible with moderate evidence in each of three datasets corresponding to two independent observations at different wavelengths, with three independent retrieval codes, has been sufficient to achieve this aim, but different works may require different specific choices to be made. We do note however that in this work we have only considered individual contributions from single molecules, and that a more exhaustive analysis in the future could consider combinations of multiple species that might be more favoured than any species individually. Furthermore, as discussed in Section \ref{sec:maximal}, while the canonical model retrievals are helpful in finding a few promising candidates, these candidates remain degenerate with each other, and, to a lesser extent, with other species. More observations are needed to be able to break this degeneracy. We show the spectra corresponding to the median results from the \texttt{POSEIDON} retrievals on the MIRI \texttt{JExoRES} dataset for DMS and methacrylonitrile in Figure \ref{fig:spectral_fits}.

\subsection{Estimated abundances}
As shown in Table \ref{tab:abundances}, for most species for which moderate evidence is found in the mid-infrared, we infer potential volume mixing ratios from the MIRI data between $\sim$10$^{-3}-$10$^{-4}$, consistent between the \texttt{JExoRES} and \texttt{JexoPipe} data instances. These values are similar to the abundances found for DMS and/or DMDS by \citetalias{madhusudhan_new_2025}. For the species which are also inferred with moderate evidence in the near-infrared with no offsets between detectors, we find overall lower ($\sim$10$^{-5}$) median abundances, but still consistent at 1$\sigma$ with the mid-infrared results. This, too, is consistent with what was found for DMS in \citet{madhusudhan_carbon-bearing_2023}. Species for which no evidence is found in the near infrared instead are found to have $2 \sigma$ abundance upper limits at $\sim$10$^{-3}-$10$^{-4}$, similar to the other non-detections reported in \citet{madhusudhan_carbon-bearing_2023}. We note that, in most of these cases, the 2$\sigma$ upper limits in the near-infrared are consistent with the median retrieved abundances in the mid-infrared. This implies that their non-detection in the near-infrared is not grounds to dismiss the possibility that they may be present in the atmosphere of K2-18~b. This caveat may also apply to other species, which, while potentially present, may lack features strong enough to allow their inference with the present data quality. This possibility is further discussed in Section \ref{sec:discussion}. Posterior distributions for select species are shown in Figure \ref{fig:posteriors}.  

\section{Physical plausibility of results}
\label{Physical plausibility of results section}

Any inference of a chemical signature on an exoplanet needs to be assessed for its physical plausibility, in the context of the environment  in which it is potentially inferred.
Such a process requires an assessment of its chemical properties, as well as photochemical models that simulate atmospheric composition under a range of plausible conditions, taking into account uncertainties in stellar spectra, chemical reaction networks, transport, and albedo \citep{Cooke2024Photo}. When observed spectral features are attributed to molecules whose predicted abundances are orders of magnitude different from plausible theoretical expectations, this raises questions about the accuracy of either the observations or the models, or both. Such discrepancies could arise from retrieval degeneracies \citep[e.g.,][]{line_influence_2016, welbanks_degeneracies_2019, welbanks_atmospheric_2022, lueber_information_2024, novais_parameter_2025} and unknown opacity sources \citep[e.g.,][]{niraula_impending_2022, niraula_origin_2023}, but they may also indicate that current chemical networks are incomplete, especially in parameter-space regimes where reaction pathways are missing or unconstrained. 

In this section, we discuss a number of promising candidate species obtained. In particular, we begin by discussing DMS, which is the only species for which moderate model preference is found across all the MIRI and NIR 0-offset retrievals. We then also discuss methacrylonitrile and allyl chloride, for which the $\ln B =2.0$ threshold for moderate preference is also reached in all our MIRI retrievals, as well as in the NIR with at least one retrieval code, for the no offset case. Then, we consider three hydrocarbons that are in common with \citetalias{welbanks_challenges_2026}, who found them with significances $\ln B \geq 2.5$ with both of the MIRI datasets, \texttt{JExoRES} and \texttt{JexoPipe}. We summarize our results for DMS, methacrylonitrile and allyl chloride in Table \ref{molNIR}, and for the other three hydrocarbons in Table \ref{molMIRI}.

\subsection{Dimethyl Sulfide}

The most promising species found in this work is dimethyl sulfide, (CH$_3$)$_2$S. 
This species is the most abundant biologically produced sulfur compound emitted to Earth's atmosphere \citep{1999JGR...104.8113S} with atmospheric mixing ratios of up to $\sim$1 ppbv, usually fluctuating between concentrations of 1--100 pptv \citep{2016ACP....16.6665M, yan2023high}. DMS is mainly produced as a byproduct of dimethylsulfoniopropionate (DMSP), which is generated by marine phytoplankton, including bacteria and algae \citep{liss1993production, ledyard1994dimethylsulfide, groene1995biogenic}, although it can be produced without DMSP as a precursor \citep{visscher2003dimethyl, carrion2015novel}. 

DMS does have astrophysical sources, as it has been observed in a comet \citep{Hanni2024} and in the interstellar medium, at abundances relative to hydrogen of $\sim1\times10^{-10}$ \citep{2025ApJ...980L..37S}. DMS is produced photochemically in laboratory experiments up to abundances of 0.6 ppmv when CH$_4$ and H$_2$S are present without CO$_2$, with the DMS abundance reducing to 0.05 ppmv as CO$_2$ is increased to 0.1\% by volume \citep{Reed2024}. There is, however, no evidence for H$_2$S in the atmosphere of K2-18 b (\citetalias{welbanks_challenges_2026, madhusudhan_new_2025}; \citealp{hu_water-rich_2025}), and there is evidence for CO$_2$ existing alongside CH$_4$ \citep{madhusudhan_carbon-bearing_2023, hu_water-rich_2025}. Consequently, as cometary delivery of plentiful DMS to be sustained for billions of years is unfeasible \citepalias{madhusudhan_new_2025}, and the interstellar medium generates negligible abundances \citep{2025ApJ...980L..37S}, 0.05 ppmv of DMS is the current known upper limit on abiotic pathways for creating DMS in an environment similar to the conditions in the atmosphere of K2-18~b. This value is $\sim 4$ dex lower than the median abundances found for DMS in \citetalias{madhusudhan_new_2025}, and $\sim 1$ dex lower than that found with the less constraining near-infrared observations in the canonical model of \citet{madhusudhan_carbon-bearing_2023}. An alternative abiotic DMS pathway has been suggested by \cite{hu_water-rich_2025}, but this requires a deep atmosphere and adequate H$_2$S, which is not detected in K2-18~b. The lack of NH$_3$ further argues against a deep atmosphere. 

Determining whether DMS (if its presence is confirmed) emanates from a biotic source will require three criteria to be met: a stronger constraint on the atmospheric DMS abundance; an inference on whether K2-18~b is a hycean world or if it has a deep atmosphere with no liquid water ocean (which \citealp{hu_water-rich_2025} have further shown to be unlikely); and ruling out abiotic sources for the inferred abundance in the given context.

\subsection{Methacrylonitrile and allyl chloride}

As discussed in Section \ref{sec:results}, the species coming closest to DMS in our analysis is methacrylonitrile (C$_4$H$_5$N, also known as methylacrylonitrile). This is also a trace chemical on Earth with no significant abundance in the atmosphere. It has been found to be toxic to mammals \citep{pozzani1968mammalian, farooqui1991toxicology}. Its industrial use includes the production of polymers and as a chemical intermediary in the synthesis of other compounds \citep{farooqui1991toxicology, grassie1956polymerization}. While it has also not been observed in any other planetary atmosphere, it has recently been located in a molecular cloud at low abundances of $10^{-11}$ relative to H$_2$ \citep{2022A&A...663L...5C}. 

Lastly, the only other species resulting in our $\ln B \geq 2.0$ threshold being met in the NIR (no offset case) with at least one of the retrieval codes we consider is allyl chloride. This molecule (C$_3$H$_5$Cl, also known as 3-Chloroprop-1-ene) is a neurotoxin which can be produced from chlorination of propylene \citep{turja2011risk} and is used in chemical synthesis \citep{berger2005human, wang2014efficient}. It is relatively insoluble in water and, whilst it is a liquid at room temperature and pressure, it is volatile and could exist as a vapor \citep{berger2005human} in the atmosphere of K2-18 b. Its abundance in Earth’s atmosphere has not been reported globally and it has not been located in astrophysical sources.

\begin{table*}[t!]
\centering
\small
\caption{Description of two of the most promising molecules across all datasets and retrievals. This table summarises their abundances and primary sources on Earth and elsewhere in the universe (whether that location is a planet, a comet, or the interstellar medium, etc.). Note that ppmv, ppbv, and pptv, mean parts per million, parts per billion, and parts per trillion by volume, respectively.  References: \textbf{(a)} \cite{2016ACP....16.6665M, yan2023high, liss1993production, groene1995biogenic, Hanni2024, 2025ApJ...980L..37S} \textbf{(b)} \cite{grassie1956polymerization, farooqui1991toxicology, 2022A&A...663L...5C} \textbf{(c)} \cite{berger2005human, wang2014efficient}} 

\begin{tabular}{p{3cm}p{2.8cm}p{3cm}p{2.5cm}p{4cm}p{0.6cm}}

\toprule
Molecule & Earth abundance & Source (Earth) & Max abundance elsewhere & Source (elsewhere) & Ref \\ 
\midrule
DMS (CH$_3$)$_2$S & $\lesssim 1$ ppbv & Algae, marine organisms & $\approx 100$ pptv & Cometary matter, interstellar medium & a \\
Methacrylonitrile (C$_4$H$_5$N) & Unknown & Chemical synthesis, polymer production & $\approx 10$ pptv & Interstellar molecular cloud & b \\
Allyl chloride (C$_3$H$_5$Cl) & Unknown & Chemical synthesis & Unknown & Unknown & c \\

\bottomrule
\label{molNIR}
\end{tabular}
\end{table*}

\subsection{Other candidates}
\label{sec:plausibility-others}
We now consider the plausibility of other candidate molecules. Specifically, we focus on the three hydrocarbons that are in common with those found in \citetalias{welbanks_challenges_2026} with preference $\ln B \geq 2.5$ in both MIRI datasets. All these molecules were found by \citetalias{welbanks_challenges_2026} to result in $\ln B$ between 2.7-3.4 with both \texttt{JExoRES} and \texttt{JexoPipe} data, comparable to DMS, similar to the present work. These are: propyne (C$_3$H$_4$), cyclohexane (C$_6$H$_{12}$), and butane (C$_4$H$_{10}$). We summarise the known abundances of these three molecules on Earth or in  astrophysical environments in Table~\ref{molMIRI}. 

Propyne (also known as methylacetylene) is not present in Earth's atmosphere at significant abundances. It is emitted by some vehicles \citep{kim2006emission, liu2008source} and has been observed in some areas at concentrations $\lesssim 20$ parts per trillion \citep[][]{2008ACP.....8..351S}. It has been found in the atmospheres of all solar system giant planets \citep{1997A&A...321L..13D, fouchet2000jupiter, Meadows2006Neptune, 2006Icar..184..634B}, with abundances generally $\lesssim10$ ppbv, and that of Pluto at $\sim$1 ppmv \citep{2020AJ....159..274S}. A mixing ratio of 4 ppmv was measured by the Cassini Ion Neutral Mass Spectrometer (INMS) in Titan's atmosphere \citep{nixon2013detection, Waite2005Titan}. Propyne has also been observed in a prestellar core \citep{2014ApJ...795L...2V}, a protostar \citep{2018MNRAS.481.5651A}, and a protoplanetary disk \citep{2024Sci...384.1086A}, as well as the interstellar medium \citep{Snyder1973Propyne}. The abundances in these environments are lower than that in the atmosphere of Titan. Whilst propyne clearly exists in nature, we do not detect it in the near infrared in the retrievals we present here. \citetalias{welbanks_challenges_2026} describe their propyne retrieval posterior with a peak near an abundance of $10^{-3}$, indicating relatively large abundances. We find a consistent result in the mid-infrared data, finding a $\log_{10}$ volume mixing ratio of $-2.84^{+0.97}_{-1.29}$ with the \texttt{VIRA} retrievals in the MIRI \texttt{JExoRES} dataset, and $-2.91^{+1.03}_{-1.36}$ when using the \texttt{JexoPipe} data instead. 

\citetalias{welbanks_challenges_2026} employed photochemical modelling using VULCAN and Photochem and also a 1D radiative–convective–photochemical equilibrium model (their Extended Fig.~3) to demonstrate that some hydrocarbons (CH$_{4}$, C$_2$H$_{2}$, C$_2$H$_{4}$, C$_2$H$_{6}$, C$_3$H$_{4}$, C$_6$H$_{6}$) can be present in abundances greater than $1$ ppmv between 0.001--1 mbar pressure range, stating that the simulated atmosphere motivates the inclusion of the HITRAN hydrocarbons \citep{HITRAN_2022_Database} in their MIRI retrievals. 

The \citetalias{welbanks_challenges_2026} photochemical results, intended to show that hydrocarbons can be produced in temperate H$_2$-dominated atmospheres (particularly propyne at $\sim$10 ppmv), are however illustrative rather than quantitative matches to the retrieved abundances for K2-18 b. For instance, the predicted abundances for H$_2$O, CO, and NH$_3$ are factors of $\sim$110, $\sim$660, and $\sim$20 times higher, respectively, than the $2\sigma$ upper limits from the \texttt{AURA} retrievals in \citet{hu_water-rich_2025}, and similarly inconsistent with the results from other retrieval codes quoted in the same study. Additionally, the detection of H$_2$S and SO$_2$ has not been reported in any K2-18~b transits, but the \citetalias{welbanks_challenges_2026} photochemical models have such molecules at mixing ratios of $\approx5\times10^{-3}$ and $\approx 10^{-4}$, respectively. These abundances might enable H$_2$S and SO$_2$ to be detected, although it is important to state that the non-detection of a molecule does not imply its absence from the atmosphere. Photochemical modelling by \citet{huang__probing_2024} predicts propyne at mixing ratios up to $\sim 10^{-2}$. However, these models also result in abundances for both detected gases (CH$_4$ and CO$_2$) and non-detected gases (e.g., H$_2$O, NH$_3$, CO) which are inconsistent with, respectively, their retrieved abundances and their upper limits \citep{madhusudhan_carbon-bearing_2023, hu_water-rich_2025}. Thus, it remains to be seen if the abundance of propyne inferred from the mid-infrared observations could be explained by photochemistry on K2-18~b. More generally, the plausibility of any given species in the atmospheres of exoplanets should be considered in the context of the known chemical constraints and account for all the uncertainties present in observations, retrievals, and photochemical models.

Another molecule of interest is cyclohexane, which is an industrial product on Earth used as a solvent or to make nylon or pesticides \citep{dada2012production}. Its emissions into Earth's atmosphere come from the petrochemical industry \citep{liu2008source} and vehicles \citep{schauer2002measurement}. It has been identified in the essences and oils of several plants \citep{macleod1982volatile, gundidza1993chemical, macleod1988volatile}. No predictions for its presence in planetary atmospheres at observable mixing ratios have been made, but it is hypothesized to be present in the interstellar medium \citep{2012MNRAS.423.2209P}.

Butane can reach up to abundances $\approx10$ ppbv in Earth's atmosphere due to it being a component of natural gas used as fuel. It is also emitted by plants \citep{schauer2001measurement}. It has been searched for in Titan's atmosphere; however, no confirmation of its presence has yet been made \citep{2022PSJ.....3...59S}. Results from Cassini's INMS indicate that butane is present in Saturn's atmosphere at mixing ratios of $\lesssim 5$ ppmv \citep{2022JGRE..12707238S}.

\begin{table*}[t!]
\centering
\small
\caption{Description of species discussed in Section \ref{sec:plausibility-others}. This table summarises their abundances and primary sources on Earth and elsewhere in the universe (whether that location is a planet, a comet, or the interstellar medium, etc.). Note that ppmv, ppbv, and pptv, mean parts per million, parts per billion, and parts per trillion by volume, respectively. References: \textbf{(a)} \cite{2008ACP.....8..351S, nixon2013detection, fouchet2000jupiter, Waite2005Titan, Snyder1973Propyne}; \textbf{(b)} \cite{macleod1982volatile, 2012MNRAS.423.2209P}; 
\textbf{(c)} \cite{schauer2001measurement,schauer2002measurement, rossabi2018changes, 2022PSJ.....3...59S, 2022JGRE..12707238S}} 
\newcolumntype{P}[1]{>{\centering\arraybackslash}p{#1}}
\begin{tabular}{@{}P{3cm} P{2.8cm} P{3cm} P{2.5cm} P{4cm} P{0.6cm}@{}}
\toprule
Molecule & Earth abundance & Source (Earth) & Max abundance elsewhere & Source (elsewhere) & Ref \\ 
\midrule
Propyne (C$_3$H$_4$) & $\lesssim 20$ pptv & Laboratory, fuel & $\approx4$ ppmv & Atmospheric photochemistry, disk chemistry, interstelar medium & a \\
Cyclohexane (C$_6$H$_{12}$) & Unknown & Industrial, plants & Unknown & Suggested to be in interstellar medium & b \\
Butane (C$_4$H$_{10}$) & $\lesssim 10$ ppbv & Fuel, plants & $\lesssim 5$ ppmv & Atmospheric chemistry & c \\
\bottomrule
\label{molMIRI}
\end{tabular}
\end{table*}

Overall, most of the molecules discussed above have not been found to have significant abiotic sources on Earth or beyond. In a few cases, some astrophysical sources are known to produce them in small quantities but their relevance to explain significant abundances in a planetary atmosphere is not clear, as discussed in Section~\ref{sec:assess} and Section \ref{sec:discussion}. These include methacrylonitrile, propyne, butane, and dimethyl sulfide. Therefore, future theoretical studies need to investigate potential biotic and abiotic mechanisms to assess the plausibility of these molecules on K2-18~b, if robustly confirmed with future observations.

\section{Summary and Discussion} \label{sec:discussion}

The JWST observations of the transmission spectrum of K2-18~b have opened a new avenue to characterize the atmospheres of temperate sub-Neptunes. In particular, the inference of DMS and/or DMDS in K2-18~b (\citealp{madhusudhan_carbon-bearing_2023}; \citetalias{madhusudhan_new_2025}) has opened the possibility of detecting potential biosignatures in habitable-zone exoplanets. However, any robust identification of a biosignature gas requires enough observations to achieve strong statistical evidence for it, as well as comprehensive theoretical works to assess the robustness of the inferences, including identifying potential false positives. In this work, we conduct a systematic and wide-ranging search for trace molecules in the atmosphere of K2-18~b, both to identify new potential species as well as to find viable alternatives to DMS and/or DMDS reported previously. We conduct this search using atmospheric retrievals over a large model space spanning contributions from 661 molecules. To ensure robustness of the results, we conduct atmospheric retrievals with three different data instances corresponding to two independent observations across the near-infrared and mid-infrared, and validate them using three independent retrieval codes. 

Our work represents a key advancement in the search for chemical species in temperate exoplanetary atmospheres with K2-18~b serving as a benchmark. The broad spectral range and high sensitivity of JWST has led to searches of increasing numbers of molecules in the atmosphere of K2-18~b. The molecular space spanned 11 species in \citet{madhusudhan_carbon-bearing_2023}, 20 in \citet{madhusudhan_new_2025} and 92 (one-by-one) in \citet{welbanks_challenges_2026}, while considering a specific wavelength range and retrieval code each. The present work spans the full range of data available, and it uses a substantially enhanced set of 661 molecules, considered individually. 
As such, our results mark an important step toward comprehensive and agnostic searches for chemical signatures in exoplanetary atmospheres.

Across the 661 molecules considered, we find only one molecule, DMS, that reach Bayes factors compatible with moderate evidence ($\ln B \geq 2.0$) when all data and retrieval codes are considered, and when uncertainties in the estimation of Bayes factors are taken into account. We find that eight molecules may reach moderate evidence levels in both available MIRI datasets obtained using two different pipelines, and with all three retrieval codes used in this work. We tested whether any of these molecules also had evidence in the near-infrared (1-5 \textmu m) transmission spectrum reported by \cite{madhusudhan_carbon-bearing_2023}. Of these, only one, dimethyl sulfide, reaches model preference values consistent with moderate evidence ($\ln B \geq 2.0$) with all three retrieval codes if no offsets are considered between NIRISS and NIRSpec; one more (methacrylonitrile) does so with \texttt{POSEIDON} and \texttt{VIRA}, but not with \texttt{pRT}. No individual species reached the moderate evidence or weak evidence threshold when allowing for an offset between the two instruments, although weak evidence is found, with \texttt{POSEIDON}, for a combination of DMS and methacrylonitrile.

Besides the molecules noted above, we find that several other molecules could explain at least some of the data considered. For example, 
propyne leads to significant model preferences with the MIRI data but not the NIR data. Secondly, we note that most of the retrievals explored in this work consider the contributions of 661 molecules individually, in addition to CH$_4$ and CO$_2$, which were inferred through the NIR observations, although not through the MIRI ones, due to the low prominence of their spectral features in the instrument's spectral range (see Figure \ref{fig:contributions}). However, future studies could explore combinations of these molecules that could provide comparable or better fits to the data compared to the present results. We note that, following the standards in the field, we rely on Bayesian analysis to assess the quality of fits, as frequentist metrics (such as $\chi^2$) suffer from a number of flaws which make them unsuitable in the context of atmospheric retrievals and hypothesis testing \citep{welbanks_application_2023}.

Consideration of maximal models in this work has shown up to moderate evidence ($\ln B = 3.2$) for excess absorption beyond standard CNO species (CH$_4$, CO$_2$, H$_2$O, CO, NH$_3$,  HCN) across the considered spectral range. However, the contribution of the individual complex species which were revealed by the canonical models to be preferred remained degenerate. 
We highlight that it is also possible that some of the molecules explored may have strong features in only one of the two wavelength ranges, near-infrared or mid-infrared, and hence may not be detectable in the other. Further observations could provide more stringent constraints on all these molecules, and lift degeneracies between them. 

Our results have important implications for the possibility of DMS in the atmosphere of K2-18~b. Previous studies reported moderate evidence for DMS with a potential degeneracy with DMDS. The present study confirms that the previously reported evidence for DMS (\citealp{madhusudhan_carbon-bearing_2023}, \citetalias{madhusudhan_new_2025}) is independent of the retrieval code used and persists across the datasets considered. Furthermore, of the promising molecules discussed above, DMS is the only one for which production mechanisms have been theorised which can explain its retrieved abundance on K2-18~b \citep{madhusudhan_habitability_2021, Tsai_2024_Sulfur}. A close second in terms of overall model preference, methacrylonitrile is also biogenic on Earth, but present in negligible amounts in the atmosphere. This molecule has not been identified in any other planet, and has no known significant abiotic pathways that can explain the retrieved abundances if present on K2-18~b. In addition, it is more complex than DMS and therefore harder to synthesize. If its presence is confirmed with future observations, theoretical studies would need to investigate its possible biotic or abiotic formation mechanisms on K2-18~b. 

\subsection{Towards systematic molecular searches}

In this work, we report an early protocol for systematic and agnostic searches of molecular species in exoplanetary atmospheres, following initial steps in \citet{madhusudhan_new_2025} and \citet{welbanks_challenges_2026}. We considered nearly the full inventory of available cross-sections in HITRAN \citep{HITRAN_2022_Database}, and assessed whether the available data provided any support for each species considered individually. Future work in this direction could explore any possible evidence for combinations of species beyond those considered in current and previous works. 

Following \citetalias{madhusudhan_new_2025}, our work demonstrates the effectiveness of using atmospheric retrievals and model preference values on a wide array of canonical models as a practical metric to restrict a very wide initial pool of species to a short list of likely candidates. In principle, an agnostic maximal model, including all considered species at once, could also be constructed, and the nested model comparison approach as described in \citet{benneke_how_2013} could be followed. However, at the present time it is computationally prohibitive to build an exhaustive maximal model. Any choice of maximal model would thus inevitably be somewhat arbitrary, unless supported by other theoretical and empirical considerations. 

Nevertheless, further work on the canonical model approach we present should also explore in greater detail its statistical aspects. For example, two opposite effects remain unaddressed. First, our work, as an early exploratory search, considers hundreds of possible species to explain the signs of excess absorption in the spectra of K2-18~b, and arrives at a ranking based on the Bayesian preferences obtained across our workflow. While this allows for a more agnostic analysis, it also introduces the multiple comparisons problem, i.e., the fact that, when a large number of hypotheses is considered to explain a given dataset, it may be expected that some result in ``statistically significant" results due to purely stochastic effects \citep[e.g.,][]{benjamini_controlling_1995, benjamini_discovering_2010}. This is a greater concern in frequentist analyses, where the statistical significance of a result is directly determined by its p-value \citep[e.g.,][]{wasserstein_asa_2016}. However, it may also be of concern in Bayesian comparisons, and hence require greater caution in the interpretation of the Bayes factors obtained for any given retrieval exploration of a dataset. Conversely, it is a general principle that the probability of independent events occurring jointly is always lower than the probability of each event individually. Thus, the fact that notable moderate preference is obtained for the same species on K2-18~b across three separate datasets corresponding to two independent observations at different wavelengths, with three different retrieval codes, makes it more likely that this result is due to a real signal, rather than random noise in the data. Future work should be carried out to accurately quantify these two effects, in order to arrive at a better interpretation of the quantitative results we present in this work.

\subsection{The importance of assessing physical plausibility}

An agnostic exploration carried out through retrievals must also be accompanied by considerations on the physical plausibility of results in the relevant context. For example, for any inferred species, potential production mechanisms -- whether photochemical, thermochemical, geological or even biological -- should be assessed, while remaining open to the possibility of unknown mechanisms existing. Furthermore, the presence in detectable amounts of species with significant sinks -- such as solubility for methacrylonitryle, or photochemistry for DMS \citep[e.g.,][]{Tsai_2024_Sulfur} -- may require constant re-generation mechanisms. 

Due consideration should also be given to the plausibility of any inferred species being detectable at the thermodynamic conditions inferred for the atmosphere. For example, dimethyl sulfide, cyclohexane, and methacrylonitrile are liquids at room temperature and pressure on Earth \citep{lide2007handbook}, although some of them readily vaporise under certain conditions. Dimethyl sulfide has a lower boiling point and a higher vapor pressure when compared to methacrylonitrile, meaning that it is less likely to condense out in  a temperate atmosphere. Observational constraints on the atmosphere's $P$-$T$ profile, together with theoretical studies on the phase these molecules may be expected to be found in at different thermodynamic conditions, may help in assessing the plausibility of their presence in the atmosphere of K2-18~b and of other temperate sub-Neptunes. It is however essential to be always aware of the possibility that unknown, exotic mechanisms may be at play in exoplanetary atmospheres, and thus caution should always be exercised when rejecting an empirically favoured solution solely on the basis of present theoretical models. 

\subsection{Future directions}

Future work in this direction should involve both observational and theoretical efforts to increase the robustness of the findings and to assess their physical plausibility. More observations will be able to increase the statistical significance of the present inferences and break potential degeneracies between the different molecules. Theoretical work is needed on multiple fronts. Further exploration of promising combinations of molecules may provide better alternatives to the present molecular inferences and also provide additional insights into the atmospheric composition of K2-18~b. While we consider the MIRI and NIR datasets as separate, future work may consider joint retrievals on them. This has already been attempted in, e.g., \citet{luque_insufficient_2025} and \citet{hu_water-rich_2025}. However, the lack of a robust prescription to account for the very different resolution and SNR of the MIRI and the NIR instruments has led to negligible differences in the retrieval results due to the MIRI data in such combined retrievals. Work is required to arrive at a well-motivated and robust prescription to account for this effect in atmospheric retrievals. At the same time, our work has highlighted the acute need for extensive molecular cross section data, both for new molecules as well as for a wide range of conditions, including different pressures, temperatures, and H$_2$ broadening for available cross sections. Finally, theoretical work is also needed to assess the physical plausibility of any inferred molecule given the astrophysical context of the exoplanet concerned.

Our work highlights the requirement for a comprehensive assessment of all available evidence linked to the physical plausibility of specific molecules being present in a planetary atmosphere, including potential biosignatures. At the same time, it demonstrates the capability of atmospheric retrievals, when used with the appropriate robustness checks, to rise to the challenge provided by the JWST datasets, and dramatically restrict the number of species whose presence may be inferred from the data, even when operating on an agnostic basis. The onset of JWST transmission observations of K2-18~b across the near- and mid-infrared wavelengths has enabled an exciting new phase of discovery in the temperate sub-Neptune regime. Further confirmation of DMS, or of other molecules considered herein, could have wide-ranging implications, from potential biosignatures to exotic abiotic chemistry.

\FloatBarrier

\vspace{4mm}
\textit{Acknowledgements:} 
N.M. and L.P.C. acknowledge support from the Science \& Technologies Facilities Council (STFC) toward the PhD studies of L.P.C. (UKRI grant No. 2886925). N.M., S.C., and G.J.C. acknowledge support from the UK Research and Innovation (UKRI) Frontier grant (grant No. EP/X025179/1; PI: N. Madhusudhan). N.M. and M.B. acknowledge support from the UK Research and Innovation (UKRI) Frontier grant (grant No. EP/X025179/1; PI: N. Madhusudhan) toward the PhD studies of M.B. N.M. and G.J.C. acknowledge support from the Leverhulme Centre for Life in the Universe. The authors would like to thank M\aa ns Holmberg and Julianne I. Moses for helpful discussions. 

\bibliography{references}{}
\bibliographystyle{aasjournal}
\FloatBarrier

\appendix

\section{Model and retrieval specifications}
\label{sec:priors}

In the below Tables \ref{tab:other_settings} and \ref{tab:all_priors}, we report our retrieval and model settings, as well as the prior distributions and descriptions for the free parameters included in our retrievals.

\begin{table}[h]
\centering
\begin{tabular}{l|c|c|c|c}
Parameter             & \texttt{POSEIDON}                                                                                                      & \texttt{VIRA}                                                                                                          & \texttt{pRT}                                                                    & Description                                                                                                          \\ \hline
$P$ grid              & \begin{tabular}[c]{@{}c@{}}10-10$^{-6}$bar, \\ 100 layers\end{tabular}                                               & \begin{tabular}[c]{@{}c@{}}10-10$^{-7}$bar, \\ 100 layers\end{tabular}                                               & \begin{tabular}[c]{@{}c@{}}10-10$^{-6}$bar, \\ 100 layers\end{tabular}        & \begin{tabular}[c]{@{}c@{}}Pressure grid \\ for the atmosphere\end{tabular}                                          \\ \hline

$\lambda$ array (MIR) & \begin{tabular}[c]{@{}c@{}}6000 points, \\ 4.8-12.0$\mu$m, \\ equal spacing in $\lambda$\end{tabular}                & \begin{tabular}[c]{@{}c@{}}6000 points, \\ 4.8-12.0$\mu$m, \\ equal spacing in $\lambda$\end{tabular}                & \begin{tabular}[c]{@{}c@{}}Constant $R =1000$, \\ 4.8-12.0$\mu$m\end{tabular} & \begin{tabular}[c]{@{}c@{}}Wavelength array \\ for MIRI retrievals\end{tabular}                                      \\ \hline
$\lambda$ array (NIR) & \begin{tabular}[c]{@{}c@{}}25000 points,\\ 0.5-5.2$\mu$m,\\ equal spacing in $\lambda$\end{tabular}                  & \begin{tabular}[c]{@{}c@{}}22000 points,\\ 0.45-5.5$\mu$m,\\ equal spacing in $\lambda$\end{tabular}                 & \begin{tabular}[c]{@{}c@{}}Constant $R =1000$, \\ 0.5-5.2$\mu$m\end{tabular}  & \begin{tabular}[c]{@{}c@{}}Wavelength array \\ for NIR retrievals\end{tabular}                                       \\ \hline

R(10$\mu$m)          & \multicolumn{1}{l|}{\begin{tabular}[c]{@{}l@{}}R(10$\mu$m) = 8332 \\ $ 170 \times \rm{R_{data}}(10 \mu m)$\end{tabular}} & \multicolumn{1}{l|}{\begin{tabular}[c]{@{}l@{}}R(10$\mu$m) = 8332 \\ $ 170 \times \rm{R_{data}}(10 \mu m)$\end{tabular}} & \multicolumn{1}{l|}{Correlated-K}                                             & \multicolumn{1}{l}{\begin{tabular}[c]{@{}l@{}}Resolution at 10\textmu m,\\ absolute \& relative\end{tabular}} \\ \hline

R(3$\mu$m)             & \multicolumn{1}{l|}{\begin{tabular}[c]{@{}l@{}} R(3$\mu$m) = 15957 \\ $3.6 \times \rm{R_{data}}(3 \mu m)$\end{tabular}}    & \multicolumn{1}{l|}{\begin{tabular}[c]{@{}l@{}}R(3$\mu$m) = 13069 \\ $2.9 \times \rm{R_{data}}(3 \mu m)$\end{tabular}} & \multicolumn{1}{l|}{Correlated-K}                                             & \multicolumn{1}{l}{\begin{tabular}[c]{@{}l@{}}Resolution at 3\textmu m,\\ absolute \& relative\end{tabular}} \\ \hline

Live points           & 2000                                                                                                                 & \begin{tabular}[c]{@{}c@{}}2000 for MIRI \\  1000 for NIR\end{tabular}                                              & 500                                                                           & \begin{tabular}[c]{@{}c@{}}Number of live \\ points for MultiNest\end{tabular}                                       \\ \hline

\begin{tabular}[c]{@{}c@{}}Sampling \\ efficiency\end{tabular}  & 0.3                                                                                                                  & 0.3                                                                                                                  & 0.3                                                                           & \begin{tabular}[c]{@{}c@{}}Sampling efficiency \\ for MultiNest\end{tabular}    \\ \hline                                    
\end{tabular}
\caption{Values of the retrieval settings and model specifications employed across this study. In \texttt{VIRA}, supersampling is used to sample the cross sections at a higher resolution (when available) than the grid resolution in the forward model. In \texttt{pRT}, correlated-K opacities are used. We also use supersampling in the NIR with \texttt{POSEIDON}; we considered the same in the MIR as well but saw no significant effect. Depending on the species, the native resolution of opacities before supersampling is up to R $\sim 3.3 \times 10^5$ at 3\textmu m and R $\sim 10^5$ at 10\textmu m.} 
\label{tab:other_settings}
\end{table}

\begin{table}[h]
\centering
\begin{tabular}{l|l|l}
\multicolumn{1}{c|}{Parameter}  &Priors                                                          & Description                                            \\  \hline
$\log({\rm X})$                       & $\mathcal{U}(-12, -0.3)$   & Volume mixing ratio of each molecule (\texttt{VIRA and \texttt{POSEIDON})}\\
$\log({\rm X_{\rm mass}})$                       & $\mathcal{U}(-11, -0.1)$   & Mass mixing ratio of each molecule (\texttt{pRT)}\\
$T_0$/K                         & $\mathcal{U}(100, 500)$  & Temperature at the top of the atmosphere                      \\
$\alpha_1$/K$^{-\frac{1}{2}}$   & $\mathcal{U}(0.02, 2.00)$ & $P-T$ profile curvature                                \\
$\alpha_2$/K$^{-\frac{1}{2}}$   & $\mathcal{U}(0.02, 2.00)$ & $P-T$ profile curvature                                \\

$\log(P_1 / {\rm bar})$         & $\mathcal{U}(-6, 0)$     & $P-T$ profile region limit                             \\
$\log(P_2 / {\rm bar})$         & $\mathcal{U}(-6, 0)$     & $P-T$ profile region limit                             \\
$\log(P_3 / {\rm bar})$         & $\mathcal{U}(-2, 0)$     & $P-T$ profile region limit                             \\
$\log(P_{\rm ref} / {\rm bar})$ & $\mathcal{U}(-6, 0)$     & Reference pressure, at $R_{\rm p}= 2.61 R_\oplus$       \\
$\log(a)$                       & $\mathcal{U}(-4, 10)$    & Rayleigh enhancement factor                            \\
$\log(\gamma)$                  & $\mathcal{U}(-20, 2)$    & Scattering slope     \\
$\log(P_{\rm c} / {\rm bar})$   & $\mathcal{U}(-6, 0)$    & Cloud-top pressure                                     \\
$\phi$                          & $\mathcal{U}(0, 1)$     & Cloud/haze coverage factor         \\
$\delta /{\rm ppm}$  & $\mathcal{U}(-100, 100)$        & Offset between NIRISS and NIRSpec detectors, when applicable         \\
\end{tabular}
\caption{Form and range of the Bayesian prior distributions for the free parameters in all of the atmospheric retrievals we carry out in this work. 
}
\label{tab:all_priors}
\end{table}

\FloatBarrier
\newpage
\section{Results from MIRI retrievals}
\label{sec:longtable}
\input{appendix_table}

Table \ref{tab:all-molecules} includes the Bayes factors obtained from MIRI canonical retrievals with \texttt{POSEIDON} for all species resulting in $\ln B \ge 1.15$ when using at least one of \texttt{JexoPipe} or \texttt{JExoRES} data. We choose $\ln B \ge 1.15$ as this corresponds to the threshold for ``substantial'' evidence in \citet{jeffreys_theory_1948} original scale, and is well within the ``weak evidence'' and ``positive evidence'' categories in respectively \citet{trotta_bayes_2008} and \citet{kass_bayes_1995}. After the table, we list all remaining species, for which we ran canonical model retrievals on both MIRI dataset, but which resulted in $\ln B < 1.15$ in both pipelines.

\subsection{Remaining species}
\label{sec:remaining}
Bromine nitrate; Boron trifluoride; 3-methylpentane; Propylene; Cyclopropane; 1-butyne; Cycloheptane; Methyltrichlorosilane; 1-pentene; Acetylene; Hydrogen cyanide; Diethyl sulfate; 1,2-epoxybutane; 2-methyl-1-pentene; 2-vinylpyridine; Cis-2-pentene; Sulfuryl fluoride; Propargyl chloride; Hexane; Cis-4-methyl-2-pentene; Ethylene; Benzene; Dimethyl sulfate; 3-methyl-1-butene; Propylene oxide; 2-methyl-1-butene; 1-heptene; Cyanogen chloride; Chloromethane; Hexanal; 3-methylhexane; 1-hexene; 1-butene; Isophorone; Heptane; Allyl bromide; 1-octene; Acrylonitrile; 1-chloro-2-methylpropane; Cyclodecane; 1,3-dichloropropane; Water; Decane; Butyraldehyde; phosphorus monosulfide (PS); Dimethyl sulfoxide; Methylium; 1,2-dichloroethane; 1-propanethiol; Undecane; Octane; Pentyl nitrate; Nonane; Styrene; 1-nonene; N,n-diethylformamide; 1-decene; 1,2,3-trimethylbenzene; 1,1,1-trifluoroethane; Isoprene; Sec-butylbenzene; 1-undecene; Propionitrile; Propane; Limonene oxide; 4-methyl-1-pentene; 4-ethyltoluene; (-)-beta-pinene; 1,3-butadiene; Diisobutylene; Isobutyraldehyde; 3-carene; Magnesium monohydride; Cis-1,2-dichloroethylene; HFC-272ca; Tridecane; Valeraldehyde; 2-methylstyrene; Thiophene; Scandium monohydride; 2,2,4-trimethylpentane; 2,3-dimethylbutane; Vinyl chloride; 2-methylpentane; 3-methylbutanal; Lithium hydride; 1,2,3,5-tetramethylbenzene; 3,4-dichloro-1-butene; 2-methyl-2-butene; Methyl vinyl ketone; Vanadium(ii) oxide; 1,2,3,4-tetramethylbenzene; Pentadecane; Tert-butylbenzene; Epichlorohydrin; 2,5-dimethylfuran; Geraniol; But-2-ynenitrile; Allyl isothiocyanate; 2-methylaziridine; 2-methyl-2-pentene; 2,4,4-trimethyl-2-pentene; Ethylbenzene; 1,2-dichloropropane; Propyl acetate; Isobutyl acetate; Limonene; Isobutane; Titanium monoxide; 4-methylpyridine; Amyl acetate; Mesitylene; Methyl acrylate; 1,2,3-trichloropropane; Cumene; Hexyl acetate; 1,1,1,2-tetrachloroethane; 1,4-dichlorobenzene; Thiirane; O-xylene; Aluminum monohydride; 1-ethyl-2-methylbenzene; Glycolaldehyde; Chlorocyclohexane; Propionaldehyde; Diketene; Butyl acetate; Carbon tetrafluoride; P-xylene; Cyclooctane; Carbon monosulfide; D-limonene; Silicon monohydride; Zirconium oxide (ZrO); Imidogen; Aluminum monoxide; Germane; 1,1,1,3,3,3-hexafluoropropane; Cycloheptene; 2-methylpyridine; Bromobenzene; Titanium; Methylglyoxal; Lithium; Propargyl alcohol; Hydrogen sulfide; Hexafluorobenzene; Cyanoacetylene; Toluene; Propylbenzene; Sodium; N,n-dimethylformamide; Hydrofluoric acid; 2,4-dimethylpentane; Isobutyronitrile; Atomic oxygen; Scandium; Ammonia; Cesium; Barium; (Z)-HFC-1234ze; Butyl isocyanate; Methyl acetate; Trans-1,3-dichloropropene; Myrcene; Barium cation; OH; Styrene oxide; Benzyl alcohol; Nitric oxide; OH+; Ethyl acrylate; Hydrochloric acid; Chromium; 4-chlorotoluene; Carbon monoxide; Magnesium; Ethynid-2-yl; Furfural; Sodium hydroxide; Iron; Methanesulfonyl chloride; Nickel carbonyl; Magnesium cation; Iron cation; Acetonitrile; Cyclohexanethiol; 2-carene; (-)-alpha-pinene; Calcium; MgO; H3+; Aluminum; Methylidyne radical; Vanadium; Iron monohydride; Calcium cation; 4-vinylcyclohexene; Titanium cation; Lanthanum monoxide; hydrosulfide (SH); Rubidium; Isoamyl acetate; Potassium; Chloroform; Manganese; Carbon disulfide; Potassium chloride; Chromium(I) hydride; Vanadium cation; Chloroacetone; Methanethiol; Titanium monohydride; 1-fluorohexane; 1,2,3,4-tetrahydronaphthalene; Nickel; Methyl isocyanate; M-xylene; Propylene sulfide; phosphorous nitride (PN); Beryllium monohydride; Tetrahydrothiophene; Hydroxyacetone; 2-propanethiol; Trichloroacetyl chloride; Hexadecane; Sulfur hexafluoride; Allene; Nitrous oxide; Isobutyl mercaptan; Butyl acrylate; Glyoxal; Thioformaldehyde; 5-methyl-2-hexanone; Cyclohexene; Allylamine; 3-methylpyridine; Acryloyl chloride; 3-ethyltoluene; Benzonitrile; 1,1,1-trichloroethane; Cyclopentene; Silicon monoxide; Carbonyl sulfide; Methyl isothiocyanate; Ethyl acetate; 2-chloroethanol; 2-nonanone; 4-methyl-2-pentanone; Sodium monoxide radical (NaO); 5-nonanol; Naphthalene; Bromochloromethane; 3-pentanone; Titanium tetrachloride; 1-penten-3-ol; Chlorosulfonyl isocyanate; Trimethylamine; Tert-amylamine; Vinyl bromide; Octafluorocyclobutane; 1-undecanol; Potassium hydroxide; Butylamine; 2-methyl-1-propanol; Nitrosyl chloride; Calcium monohydride; 1-hexanol; Acetaldehyde; 2-hexanone; 1-nonanol; 1-butanol; 2-pentanone; 4-penten-1-ol; Allyl fluoride; 2-mercaptoethanol; 4-methyl-1-pentanol; Quinoline; 2-chlorotoluene; Allyl alcohol; Methanol; Methyl benzoate; 1,1,2-trichloroethane; Fluoroacetone; 3-methyl-2-pentanone; Trichloroethylene; Sodium hydride; 1-pentanol; Menthol; Tetrafluorosilane; Perfluoroisobutylene; 2-chloropropane; Diacetone alcohol; 2-butanone; Ethanol; Sulfur monoxide; Furfuryl alcohol; Pentane-1,5-diamine; 3,3-dimethyl-2-pentanol; Boron tribromide; 2-bromopropane; Isoamyl alcohol; Methyl methacrylate; 2-butanol; Cyclohexanone; Phosphorus monoxide; 1-heptanol; Acrolein; 1-propanol; 2-ethylhexan-1-ol; Tert-butyl acetate; 2,3-butanedione; 1,1-dimethylhydrazine; Nitrogen trifluoride; Isocyanic acid; Triethylamine; 3-pentanol; Allyl iodide; Acetic acid; 1,1,2,2-tetrachloroethane; Pyridine; Silane; Hexafluoroacetone; Diisopropylamine; Arsine; 2-pentanol; 3-chlorotoluene; Ozone; Flurothyl; Peroxyacetyl nitrate; Tetrachloroethylene; 2-hexanol; Ethylamine; Crotonaldehyde; Trans-2-butene; (e)-1,2,3,3,3-pentafluoroprop-1-ene; Benzaldehyde; Diethylamine; Thiophosphoryl chloride; 2-pentylfuran; Neopentyl alcohol; Piperidine; 1,2-dibromo-3-chloropropane; Dichlorosilane; Ethyl benzoate; Phosphorus oxychloride; 2,6-diethylaniline; 3-chloro-1,1,1-trifluoropropane; Methylamine; Hexamethylphosphoramide; 1,1,1,2,3,3,3-heptafluoropropane; Methylhydrazine; Fluoroethane; 2-iodopropane; 1,2-dichloroethylene; 3-methyl-2-butanone; Dimethylamine; Hydrazine; Tert-butanol; Sulfuryl chloride; 1,1,1,3,3-pentafluorobutane; Perchloromethyl mercaptan; 3-methylfuran; Benzyl bromide; Methyl pivalate; Ethylene glycol; Isopropylamine; 2-methylfuran; Trifluoroacetyl chloride; Isopropyl acetate; NO2; HFE-263m1; sec-Amylamine; Cis-1,3-dichloropropene; Cis-2-butene; Phosphine; Bromoform; Trifluoroacetaldehyde; 1,1,1-trichlorotrifluoroethane; Ethylenediamine; 1,2-difluoroethane; Isopropyl alcohol; 1,3-dichlorobenzene; Propylene glycol; 1,2-dichlorobenzene; Methyl acetoacetate; 2-methyl-2-propanethiol; Ethyl nitrite; 1-nitropropane; Fluoromethane; O-toluidine; Hexafluoropropene; Nitroethane; Diiodomethane; 1,2-dibromoethane; 2,3-dichloro-1-propene; Cyclohexanol; 2-methoxy-4-nitrophenol; Nitrous acid; Pentacarbonyliron; 1,1,1,2,2-pentafluoro-2-(trifluoromethoxy)ethane; Sevoflurane; Benzenethiol; Nicotine; 1,1-difluoroethane; Nitromethane; 2,2,2-trifluoroethyl methyl ether; Acetyl chloride; Trans-1,2-dichloroethylene; 1,1,1,2-tetrafluoroethane; Furan; Morpholine; 4-methylvaleric acid; Perfluoropropane; Texanol; 2-chloro-1,1,1-trifluoroethane; 2-ethoxyethyl acetate; Acetone cyanohydrin; 3,3,3-trifluoropropanal; Ethyl butyrate; Methyl nitrite; Methyl salicylate; Eucalyptol; 3-Methoxyperfluoro(2-methylpentane); Bromodichloromethane; Aniline; Diisopropyl ether; Thionyl fluoride; 2-nitropropane; Sulfur dioxide; Trichloromethanol; Methyl hexanoate; 3,3,3-trifluoropropene; 2-fluoroethanol; Trifluoromethyl trifluorovinyl ether; 1,1,1,3,3,3-hexafluoro-2-propanol; 2,3-dimethylfuran; Chlorobenzene; Butyric acid; 2,3,3,3-tetrafluoro-2-(trifluoromethyl)propanenitrile; Oxirane; Dimethylcarbamoyl chloride; Dipropyl ether; N,n-diethylaniline; 1,1,1,2,2-pentafluoropropane; Trifluoromethyl perfluoropropyl ether; Methyl propionate; 3,3,3-trifluoro-1-propanol; 4-methoxyphenol; Methyl butyrate; Vinyl acetate; Tert-butyl ethyl ether; Dichlorofluoromethane; 2-methoxy-5-nitrophenol; 1,2-dibromoethylene; Dibromomethane; 1,2-dichlorofluoroethane; Tetrahydrofuran; Tert-butyl methyl ether; Methyl 2-methylbutyrate; Dichloromethanol; Nonenedioic acid; Perfluoro-2-methyl-3-pentanone; Octanoic acid; Perfluoroisohexane; Chloromethyl ethyl ether; Methyl vinyl ether; Chlorodibromomethane; 1-methoxybutane; Chlorotrifluoromethane; Halothane; Sulfur, pentafluoro(trifluoromethyl)-; Hexanoic acid; 2-chloroethyl ethyl ether; Benzoyl chloride; Chloropentafluoroethane; Dodecafluoropentane; Tert-amyl methyl ether; Guaiacol; 1,1-dichloro-1-fluoroethane; 1,2-dichloro-1,1-difluoroethane; Ppg-2 methyl ether; Hexachlorocyclopentadiene; Acrylic acid; Nitrobenzene; 1,1,1,3,3-pentafluoropropane; Methyl isobutyrate; 2-butoxyethanol; Methyl nonafluorobutyl ether; Dinitrogen pentaoxide; 1,4-dioxane; Trichlorofluoroethylene; 2-methoxyethanol; Vinyl fluoride; Hexafluoroisobutylene; Trichlorofluoromethane; Ethyl vinyl ether; Trifluoronitrosomethane; 1,1,1,2-Tetrafluoro-2-(trifluoromethoxy)ethane; Allyl trifluoroacetate; (E)-Perfluorodecalin; HFE-245fa2; Hexachloro-1,3-butadiene; M-cresol; Methyl nonafluoro-n-butyl ether; Allyl 1,1,2,2-tetrafluoroethyl ether; Chlorofluoromethane; Chloromethyl methyl ether; Chloromethanol; Hexafluoroisopropyl methyl ether; 1,2-dichloro-1,2-difluoroethane; 1,1-dichlorotetrafluoroethane; Valeric acid; 2,6-dimethoxyphenol; Perflubutane; (E)-HCFO-1233zd; 1,2-dichloro-1,1,3,3,3-pentafluoropropane; 1,1,1-trifluoroacetone; Chloropicrin; 1,1,2,2-tetrafluoro-3-methoxypropane; Chlorotrifluoroethylene; Isobutyric acid; Fluorobenzene; 1-(ethoxy)nonafluorobutane; Ethyl formate; Propionic acid; 1h,1h,2h,2h-perfluoroheptan-1-ol; 2,2,3,3-tetrafluoro-1-propanol; 3,3,4,4,5,5,6,6,7,7,8,8,9,9,9-pentadecafluorononan-1-ol; 3-methoxyphenol; Methyl formate; HFE-356mec3; 1,1-Dihydroperfluoropropanol; Ethyl chloroformate; 2,2,3,4,4,4-hexafluoro-1-butanol; Tetrafluoroethylene; Vinylidene chloride; CFC-112; 1,1,2,3,3,4,4,4-octafluorobut-1-ene; HG-10; HFE-7500; HFE-125; HFC-32; Methoxyethane; HCFC-121; 1,1-dibromotetrafluoroethane; 1,1-dichloro-1,2-difluoroethane; Thiophosgene; 2,4-diisocyanato-1-methylbenzene; Chlorodifluoromethane; Bis(2-chloroethyl) ether; Dimethyl ether; 3,3,4,4,5,5,6,6,6-nonafluoro-1-hexene; 1,1,1,2,2,3-hexafluoropropane; 1,3-dioxolane; Bromodifluoromethane; 2,2-dichloro-1,1,1-trifluoroethane; Phenol; Diethyl ether; Dibromodifluoromethane; Ethyl 1,1,2,3,3,3-hexafluoropropyl ether; Hexafluoroethane; HFE-365mcf3; 2,2,2-trifluoroethanol; 1,1-difluorotetrachloroethane; Heptafluorobutyraldehyde; 1,2-dimethoxy-1,1,2,2-tetrafluoroethane; Dichlorodifluoromethane; HFC-43-10mee; 1,1,2-trifluoroethane; 2,3,3,3-tetrafluoropropene; 1,2-dimethoxyethane; Nonafluoropentanal; Heptafluoropropane; 2,2-difluoroethan-1-ol; 2-methyl-1,3-dioxolane; Perfluoroheptane; Propylene carbonate; Methyl chloroformate; 1,1-dichloro-2,2-difluoroethane; 1,1,1,2,3,3-hexafluoropropane; 3,3,4,4,5,5,6,6,7,7,8,8,9,9,10,10,11,11,11-Nonadecafluoroundecan-1-ol; (Z)-Perfluorodecalin; Trifluoromethyl methyl ether; Isoflurane; Trifluoroacetic anhydride; Pentafluoroethane; Paraldehyde; Dimethoxymethane; Trifluoromethane; 2-chloro-1,1,1,2-tetrafluoroethane; Hexafluoro-1,3-butadiene; 2,2,3,3,4,4,4-heptafluoro-1-butanol; (z)-1,2,3,3,3-pentafluoropropene; Dichlorotetrafluoroethane; Ethyl trifluoroacetate; Diborane; 2,2,3,3,3-pentafluoropropanal; 1,1,2-trichlorotrifluoroethane; HG-02; 3,3,4,4,4-pentafluorobut-1-ene; 1-chloro-1,1-difluoroethane; Enflurane; Perfluorotributylamine; 1,2-dichloro-1,1,2-trifluoroethane; Trifluoroacetic acid; 1h-perfluorohexane; 1,1,1,2,2,3,3,4,4-nonafluorobutane; Perfluorohexane; Acetic anhydride; Desflurane; 3,3-dichloro-1,1,1,2,2-pentafluoropropane; 1-chloro-1,1,2,2-tetrafluoroethane; Perfluoropolymethylisopropyl ether; Propane, 1,1,1,2,2,3,3-heptafluoro-3-methoxy-; Perfluoro-2-methyl-2-pentene; HG-11; HFE-347pcf2; HFE-338mec3; Vinyl trifluoroacetate; Heptafluoropropyl 1,2,2,2-tetrafluoroethyl ether; 1h,1h,2h-perfluoro-1-decene; Perfluorooctane; 1,3-dichloro-1,1,2,2,3-pentafluoropropane; (Perfluorohexyl)ethylene; Bromotrifluoromethane; 1,1,2,2-tetrafluoroethane; HG-03; Bromochlorodifluoromethane; 1,1,2,2,3-pentafluoropropane; Decamethylcyclopentasiloxane; HG-01.
\end{document}

%% file: appendix_table.tex
\begin{table*}[ht]
\centering
\caption{Comparison of log Bayes factors and significance ($\ln B$ and $\sigma$) for molecules retrieved using \texttt{POSEIDON} from JWST MIRI observations, which result in $\ln B \ge 1.15$ for at least one of \texttt{JExoRES} and \texttt{JExoPipe} reductions. The Bayes factors are for the canonical model with the relevant molecule included, versus an identical baseline model only including CH$_4$ and CO$_2$, as discussed in Section \ref{sec:methods}. The Bayesian evidence of the reference model against which Bayes factors are reported is $\ln(Z) = 212.82$ when using the \texttt{JExoRES} data and $\ln(Z) = 208.18$ with the \texttt{JExoPipe} data. Associated cross-section metadata are also included for species not originally present in the \texttt{POSEIDON} database.\label{tab:all-molecules}}
\begin{tabular}{llllllll}
\hline
Molecule & \multicolumn{2}{c}{JExoRES} & \multicolumn{2}{c}{JExoPipe} & \multicolumn{3}{c}{Cross-section metadata} \\
 & $\ln B$ & $\sigma$ & $\ln B$ & $\sigma$ & $P$ [bar] & $T$ [K] & Range [$\mu$m] \\
\hline
Chloroethane & 3.06 & 2.97 & 3.95 & 3.28 & 1.01 & 298.10 & 1.54--16.95 \\
Cyclohexane & 2.53 & 2.77 & 2.98 & 2.94 & 1.01 & 298.10 & 1.54--17.54 \\
Dichloromethane & 2.55 & 2.78 & 2.79 & 2.87 & 1.01 & 298.10 & 1.54--16.67 \\
Propyne & 2.56 & 2.78 & 2.74 & 2.85 & 1.01 & 298.10 & 1.49--17.86 \\
Bromoethane & 2.64 & 2.81 & 2.04 & 2.56 & 1.01 & 298.10 & 1.54--18.87 \\
Methacrylonitrile & 2.56 & 2.78 & 2.37 & 2.70 & 1.01 & 298.10 & 1.54--19.23 \\
2-butene (cis and trans) & 2.51 & 2.76 & 0.84 & 1.92 & 1.01 & 298.10 & 1.54--18.18 \\
Tetramethylsilane & 1.59 & 2.35 & 2.51 & 2.76 & 1.01 & 298.10 & 1.35--18.18 \\
Dimethyl sulfide & 2.22 & 2.64 & 2.51 & 2.76 & 1.01 & 298.10 & 1.54--16.67 \\
Butane & 2.45 & 2.73 & 2.14 & 2.60 & 1.01 & 298.10 & 1.54--16.67 \\
Methacryloyl chloride & 2.43 & 2.73 & 1.27 & 2.18 & 1.01 & 298.10 & 1.54--18.87 \\
Methacrolein & 2.42 & 2.72 & 1.25 & 2.17 & 1.01 & 298.10 & 1.54--16.67 \\
Dimethyl carbonate & 1.45 & 2.28 & 2.39 & 2.71 & 1.01 & 298.10 & 1.54--17.39 \\
Cyclopentane & 2.38 & 2.70 & 2.16 & 2.61 & 1.01 & 298.10 & 1.54--19.61 \\
Dichloromethylphosphine & 1.02 & 2.04 & 2.32 & 2.68 & 1.01 & 298.10 & 1.54--19.23 \\
Dimethyl disulfide & 2.30 & 2.67 & 1.91 & 2.50 & 1.01 & 298.10 & 1.54--16.67 \\
1-chloropentane & 1.22 & 2.15 & 2.17 & 2.62 & 1.01 & 298.10 & 1.54--16.95 \\
Allyl chloride & 2.14 & 2.60 & 2.05 & 2.56 & 1.01 & 298.10 & 1.54--19.05 \\
Ethanethiol & 1.36 & 2.23 & 2.08 & 2.58 & 1.01 & 298.10 & 1.54--17.09 \\
Trans-2-pentene & 2.02 & 2.55 & 1.05 & 2.05 & 1.01 & 298.10 & 1.54--18.18 \\
Diethyl sulfide & 1.77 & 2.44 & 1.96 & 2.52 & 1.01 & 298.10 & 1.54--16.67 \\
Boron trichloride & 1.93 & 2.51 & 0.86 & 1.93 & 1.01 & 298.10 & 1.54--16.67 \\
Tetradecane & 1.55 & 2.33 & 1.92 & 2.50 & 1.01 & 298.10 & 1.54--17.24 \\
Iodomethane & 1.88 & 2.49 & 1.59 & 2.35 & 1.01 & 298.10 & 1.45--20.00 \\
1-chlorobutane & 1.00 & 2.02 & 1.68 & 2.39 & 1.01 & 298.10 & 1.54--18.52 \\
Tungsten hexafluoride & 1.52 & 2.31 & 0.68 & 1.81 & 1.01 & 298.10 & 1.54--16.67 \\
Isobutylene & 1.41 & 2.25 & 1.27 & 2.18 & 1.01 & 298.10 & 1.54--16.67 \\
Chlorine nitrate & 0.21 & 1.39 & 1.39 & 2.24 & 0.10 & 297.40 & 7.58--7.94 \\
Iodoethane & 1.33 & 2.21 & 0.21 & 1.39 & 1.01 & 298.10 & 1.54--16.67 \\
2-methylbutane & 1.01 & 2.03 & 1.32 & 2.21 & 1.01 & 298.10 & 1.54--16.67 \\
Calcium oxide & 1.21 & 2.15 & 1.19 & 2.13 & --- & --- & --- \\
Neopentane & 0.61 & 1.76 & 1.19 & 2.14 & 0.14 & 293.80 & 6.45--8.13 \\
Ethane & 1.19 & 2.13 & 0.65 & 1.79 & --- & --- & --- \\
1,1-dichloroethane & 0.74 & 1.85 & 1.19 & 2.13 & 1.01 & 298.10 & 1.54--17.86 \\
2,2-dimethylbutane & 1.15 & 2.12 & 0.59 & 1.74 & 1.01 & 298.10 & 1.54--16.67 \\
Pentane & 0.89 & 1.96 & 1.15 & 2.11 & 1.01 & 298.10 & 1.54--17.54 \\
\hline
\end{tabular}
\end{table*}